\newif\iffigs\figsfalse

\iffigs \input epsf \else  \fi

\documentstyle[12pt]{article}
\catcode`\@=11
\@addtoreset{equation}{section}

\def\section{\@startsection{section}{1}{\z@}{3.5ex plus 1ex minus
   .2ex}{2.3ex plus .2ex}{\large\bf}}
\def\subsection{\@startsection{subsection}{2}{\z@}{2.3ex plus
   .2ex}{2.3ex plus .2ex}{\bf}}
\newcommand\Appendix[1]{\def\thesection{Appendix \Alph{section}}
  \section{#1} \def\thesection{\Alph{section}} }

\newlength{\pubnumber} 
\newcommand\pubblock[2]{\begin{flushright}\parbox{\pubnumber}
 {\begin{flushleft}#1\\ #2\\ hep-th/9310131\\
 \end{flushleft}}\end{flushright} }

\relax
\def\ket#1{\vert #1\rangle}
\def\IZ{{\bf Z}}

\begin{document}
\begin{titlepage}
\samepage{
\pubblock{IASSNS-HEP-93/57}{CLNS 92/1176}
\vfill
\begin{center}
{\Large \bf Tree scattering amplitudes of the\\
spin-4/3 fractional superstring I:\\
The untwisted sectors\\}
\vfill
{\large Philip C. Argyres$^1$\footnote{e-mail: argyres@guinness.ias.edu}
and S.-H. Henry Tye$^2$\\}
\vspace{.25in}
$^1${\it School of Natural Sciences, Institute for Advanced Study,
Princeton, NJ 08540}\\
$^2${\it Newman Laboratory of Nuclear Studies, Cornell University,
Ithaca, NY 14853}\\
\end{center}
\vfill
\begin{abstract}
Scattering amplitudes of the spin-4/3 fractional superstring are shown
to satisfy spurious state decoupling and cyclic symmetry (duality) at
tree-level in the string perturbation expansion.  This fractional
superstring is characterized by the spin-4/3 fractional superconformal
algebra---a parafermionic algebra studied by Zamolodchikov and Fateev
involving chiral spin-4/3 currents on the world-sheet in addition to
the stress-energy tensor.  Examples of tree scattering amplitudes are
calculated in an explicit $c=5$ representation of this fractional
superconformal algebra realized in terms of free bosons on the string
world-sheet.  The target space of this model is three-dimensional flat
Minkowski space-time with a level-2 Ka\v{c}-Moody $so(2,1)$ internal
symmetry, and has bosons and fermions in its spectrum.  Its closed
string version contains a graviton in its spectrum.  Tree-level
unitarity (i.e., the no-ghost theorem for space-time bosonic physical
states) can be shown for this model.  Since the critical central charge
of the spin-4/3 fractional superstring theory is 10, this $c=5$
representation cannot be consistent at the string loop level. The
existence of a critical fractional superstring containing a
four-dimensional space-time remains an open question.
\end{abstract}
\vfill}
\end{titlepage}

\section{\label{sI}Introduction}

String theories are characterized by the local symmetries of
two-dim\-en\-sion\-al field theories on the string world-sheet.  The
bosonic string is invariant under diffeomorphisms and local Weyl
rescalings on the world-sheet, and the superstring is characterized by
a locally supersymmetric version of these symmetries.  It is natural to
ask whether other symmetries on the world-sheet can give rise to
consistent string theories.  Since fractional-spin fields exist in
two-dimensional theories, one can imagine new local symmetries on the
world-sheet involving fractional-spin currents (replacing the spin-3/2
supercurrent of the superstring).  A proposal for a large class of new
string theories, called fractional superstrings, based on these
fractional symmetries was advanced in Ref.\ \cite{ALT}.  The critical
central charges of the fractional superstrings are smaller than
that of the ordinary superstring.  Evidence has been presented for
the existence of fractional superstrings with potentially realistic
phenomenologies in space-times of dimension four and six \cite{AT,ADT}.
This paper presents a spin-4/3 fractional superstring
model that is consistent at tree level in string perturbation theory,
and has a low-energy spectrum and scattering amplitudes describing
gravity, Yang-Mills theory, and fermions.

The basic idea behind the fractional superstring is to replace the
world-sheet supersymmetry of ordinary superstring theory with a
world-sheet ``fractional supersymmetry''.  Such a fractional
supersymmetry relates world-sheet coordinate boson fields $X^\mu$ not
to fermions but rather to fields $\epsilon^i_\mu$ of fractional
world-sheet spin $h$.  The fractional supersymmetry is generated
by a generalization of the supercurrent, a set of new chiral
currents $G^i$ \cite{ZFpf,ZFft,KMQ} whose conformal dimensions are
$1+h$.  The computationally simplest case after the ordinary
superstring is the spin-4/3 fractional superstring where $h=1/3$,
and is the subject of this paper.
The dimension-4/3 fractional supercurrents $G^\pm$ are of the form
	\begin{equation}
	 G^\pm(z)\sim\epsilon^\pm_\mu\partial X^\mu+\ldots~,
	\end{equation}
and generate, along with the stress-energy tensor $T$, the
spin-4/3 fractional superconformal algebra.  Classically, this spin-4/3
algebra is the constraint algebra arising from gauge-fixing the local
world-sheet symmetry.  Quantum mechanically, the constraints generate
physical state conditions which pick out the propagating degrees of
freedom from the larger string state space.  Although the classical
world-sheet gauge symmetry giving rise to a spin-4/3 constraint algebra
is not understood at present, we can make progress by taking the
constraint algebra itself as a starting point, and checking the
consistency of the resulting string theory by constructing unitary
scattering amplitudes for the physical states.
This approach mimics the original construction of the superstring.

By analogy with the superconformal gauge of the superstring, the
stress-energy tensor and fractional supercurrents are assumed to
generate the physical state conditions.  In particular, physical states
are taken to be annihilated by all the positive modes of $T$ and
$G^\pm$.  The physical states are thus highest-weight states of the
fractional superconformal algebra.  In Section \ref{sII} we derive the
properties of a class of highest-weight modules of this algebra---the
{\it untwisted} modules---using the techniques developed by
Zamolodchikov and Fateev \cite{ZFpf,ZFft} for studying parafermionic
algebras.  These modules are organized by a $\IZ_3$ symmetry of the
fractional superconformal algebra.  Highest-weight states with
$\IZ_3$ charge $\pm1$ are said to belong to D-modules, while those
with $\IZ_3$ charge 0 are in S-modules.  The results we derive for
these modules are independent of the choice of particular
conformal field theory representations of the spin-4/3 fractional
superconformal algebra.

We then take a first step towards showing the consistency of fractional
superstrings by defining tree scattering amplitudes and showing that
they are consistent with the assumed physical state conditions
following from the spin-4/3 fractional superconformal algebra.  In
other words, in tree scattering, physical states never scatter to
unphysical states, and null states can also be consistently decoupled
from scattering of other physical states.  This property is commonly
referred to as {\it spurious state decoupling}, and is shown in Section
\ref{sIII} in a representation-independent way.  The argument for
spurious state decoupling follows closely that used in the ``old
covariant formalism'' \cite{NST} for ordinary superstring amplitudes;
however, due to the non-linearity of the spin-4/3 fractional
superconformal algebra, an extra independent cancellation is required
for the argument to succeed compared to the ordinary superstring case.
The fact that this cancellation does occur is not trivial, and is
additional evidence for the basic consistency of fractional superstrings.

Scattering of D-module states can be written in three physically
equivalent ``pictures,'' reflecting the $\IZ_3$ symmetry of the
fractional superconformal algebra, in which the vertex operators
for scattering can be one of $W^\pm$ of conformal dimension 1/3
and $\IZ_3$ charge $\pm1$ or $V^{(+)}$ of conformal dimension 1
and $\IZ_3$ charge 0.  This is closely analogous to the two
different pictures for scattering of Neveu-Schwarz sector states in the
old covariant formalism for the ordinary superstring, in which vertex
operators can be either G-parity even dimension-1/2 operators or
G-parity odd dimension-1 operators.  Scattering of
S-module states is more problematic due to the absence of an
appropriate dimension-1 vertex operator in that sector.  In this
respect the S-module states are analogous to the Ramond sector states
of the ordinary superstring.  However, unlike the Ramond sector, the
S-module sector includes the scalar ground state of the spin-4/3
string.  {}From this point of view the S-module states are analogous to
the GSO-projected states of the Neveu-Schwarz sector, which include a
tachyon state.  Indeed, S-module states can also be shown to decouple
from tree scattering amplitudes with D-module states by a $\IZ_3$
analog of the GSO projection \cite{GSO}.

A separate issue that can be addressed at tree level is the
unitarity of scattering amplitudes.  In particular, spurious state
decoupling implies unitarity only if one can prove that the space of
physical states has non-negative norm.  This latter property is called
the {\it no-ghost theorem}.  We will not prove a no-ghost theorem in
this paper; however, such a theorem is proven in Ref.\ \cite{AKT} for
the three-dimensional model presented in Section \ref{sIV} of this paper.

The model presented in Section \ref{sIV}
is a particular conformal field theory
representation of the spin-4/3 fractional superconformal algebra with
central charge $c=5$.  It is made up of three free coordinate boson
fields $X^\mu$ on the world-sheet and a two-boson representation of the
$so(2,1)_2$ Wess-Zumino-Witten model.  This model thus has a global
three-dimensional Poincar\'{e} invariance.  The non-linear nature of
the spin-4/3 fractional superconformal algebra makes the existence of
such a representation non-trivial.  Also, the states in the model are
found to be space-time bosons or fermions, showing that the existence
of fractional-spin constraints on the world-sheet need not imply
fractional spins in space-time. The untwisted sectors of the fractional
superconformal algebra describe space-time bosonic physical states in
this representation.   Some simple states and their scattering
amplitudes are discussed in Section \ref{sIV}.  In particular, the
lowest-mass D-module states describe massless gauge fields for the open
string and a graviton for closed or heterotic-type fractional
superstrings.  Appendix A collects some useful details of the free
boson construction of the $so(2,1)_2$ conformal field theory.  Appendix
B briefly describes other known representations of the spin-4/3
fractional superconformal algebra.

Fields transforming in $so(2,1)_2$ spinor representations in the $c=5$
representation ({\it i.e.}\ as space-time fermions) appear in the
twist-sector of the $\IZ_2$ orbifold of the two-boson theory
describing the $so(2,1)_2$ current algebra.  The resulting physical
states are highest-weight states of {\it twisted} modules of the FSC
algebra.  A companion paper \cite{tree2} discusses the properties of
these modules and the spurious state decoupling argument for scattering
amplitudes involving the twist-sector highest-weight states.

The structure of the highest-weight modules of the spin-4/3 fractional
superconformal algebra ({\it i.e.}, its Ka\v{c} determinant formula)
can be used to place restrictions on the values of the central charge
and the intercepts in various sectors consistent with unitarity.  In
the bosonic and superstrings, for representations with one time-like
(space-time) dimension, a non-negative physical state space occurs up
to a maximum value of the central charge.  As one passes through this
critical value of the central charge the norm of some physical states
change sign, implying that at the critical central charge there are
extra null states.  Thus, one can check for the existence of a critical
central charge in a representation-independent way by searching for the
occurence of extra sets of zero-norm physical states.  For the spin-4/3
fractional superstring, the critical value of the central charge is
found to be $c=10$ \cite{ALT}.

One immediate consequence of this value of the critical central charge
is that the three-dimensional spin-4/3 fractional superstring model
that we present as an example in Section \ref{sIV} is {\it not} a
critical string since its central charge is $c=5$.  This fact, however,
has no significance at the level of tree scattering amplitudes.  It is
only for loop amplitudes that one expects the condition $c=10$ to
manifest itself, since it comes from an anomaly cancellation
condition.  Indeed, this is precisely what occurs in the bosonic and
superstrings, where unitary tree amplitudes exist for $c\le26$ and
$c\le15$, respectively, but loop amplitudes are only sensible at the
upper bounds of these ranges, {\it i.e.}\ at the critical central
charges.  Since three-dimensional Minkowski space-time is too small to
decribe nature, it is encouraging that the central charge of our
three-dimensional model is less than 10, allowing the possibility of a
critical spin-4/3 fractional superstring containing four-dimensional
Minkowski space-time.

This paper is organized so that the technical matter appears in Section
\ref{sII}.  Since some readers may be unfamiliar with the
considerations involved in analysing the properties of highest-weight
modules of parafermionic algebras, we have tried to make the other
parts of the paper intelligible without reading that section.  In
particular, if the reader reads only the first two paragraphs of
Section \ref{sII} and is willing to accept the results
summarized in Eqs.~(\ref{Smod}--\ref{Sdesc}) and
(\ref{Dint}--\ref{Ddesc2}), the discussion of spurious state decoupling
in tree scattering amplitudes in Section \ref{sIII} should be
self-contained.  However, we have tried to provide sufficient detail to
make the arguments of Section \ref{sII} intelligible to any reader
familiar with the basics of two-dimensional conformal field theory.
Also, the three-dimensional model given in Section \ref{sIV} provides a
concrete example of the abstract considerations of Section \ref{sII},
in which all the computations are easy to carry through since only free
scalar fields are involved.

\section{\label{sII}The spin-4/3 fractional superconformal
algebra}

The fractional currents, $G^\pm(z)$, and the energy-momentum tensor,
$T(z)$, together generate the fractional superconformal (FSC) chiral
algebra, encoded in the singular terms of their operator product
expansions (OPE):
	\begin{eqnarray}
   \label{fsca}
	T(z)T(w)&=&{1\over(z-w)^4}\left\{{c\over2}+2(z-w)^2T(w)
	+(z-w)^3\partial T(w)+\ldots\right\}~,\nonumber\\
	T(z)G^\pm(w)&=&{1\over(z-w)^2}\left\{{4\over3}G^\pm(w)
	+(z-w)\partial G^\pm(w)+\ldots\right\}~,\nonumber\\ G^+(z)
	G^+(w)&=& {\lambda^+\over(z-w)^{4/3}} \left\{
	G^-(w)+{1\over2}(z-w)\partial G^-(w)+\ldots\right\},\nonumber\\
	G^-(z) G^-(w)&=& {\lambda^-\over(z-w)^{4/3}} \left\{
	G^+(w)+{1\over2}(z-w)\partial G^+(w)+\ldots\right\},\nonumber\\
	G^+(z) G^-(w)&=& {1\over(z-w)^{8/3}} \left\{{3c\over8}+ (z-w)^2
	T(w)+\ldots\right\}.
	\end{eqnarray}
The first OPE implies that $T(z)$
obeys the conformal algebra with central charge $c$, while the second
implies that $G^\pm(z)$ are dimension-4/3 Virasoro primary fields.  The
FSC algebra was first studied by Zamolodchikov and Fateev
\cite{ZFpf,ZFft}.  The constants $\lambda^\pm$ in the $G^\pm G^\pm$
OPEs are real parameters which are definite functions of $c$.  We will
show below that associativity fixes $\lambda^+\lambda^- = (8-c)/6$.
Using a remaining freedom to rescale the $G^\pm$ currents, we choose the
conventional values of $\lambda^\pm$ to be
	\begin{eqnarray}
   \label{lamb} \lambda^+ &=& \lambda^- = \sqrt{8-c\over6}
	,\qquad{\rm for}\ c<8 ,\nonumber\\ \lambda^+ &=& -\lambda^- =
	\sqrt{c-8\over6} ,\qquad{\rm for}\ c>8 .
	\end{eqnarray}
Conformal invariance fixes all the other coefficients in (\ref{fsca}).
This algebra generates the physical state conditions for the spin-4/3
fractional string.  The holomorphic chiral algebra (\ref{fsca}) is
suitable for describing open string states and scattering.  For closed
strings one must include an antiholomorphic copy of this algebra.  We
will focus almost exclusively on the open string case, since the
generalization to closed string states and tree scattering amplitudes
is straight-forward.

An important property of the FSC algebra is its group of
automorphisms, which organizes the representation theory of its
highest-weight modules.  The order-six automorphism group $S_3$ of the
FSC algebra is generated by the transformations
	\begin{eqnarray}
   \label{autom} G^\pm &\rightarrow &\omega^{\pm1}G^\pm ,\nonumber\\
	G^\pm &\rightarrow & {\rm sign}(8-c) G^\mp ,
	\end{eqnarray}
where $\omega=e^{2\pi i/3}$ is a cube-root of unity.  We will exploit
the $\IZ_3$ subgroup of automorphisms generated by the first
transformation in (\ref{autom}) in the remainder of this section to
analyze the properties of the untwisted modules of the FSC algebras
using CFT techniques developed for parafermionic algebras
\cite{ZFpf,ZFft}.  A basis of states in the untwisted modules can be
taken to have definite $\IZ_3$ charges $q$.  Highest-weight states with
$q=0$ are said to be in an S-module, while D-modules have pairs of
highest-weight states with $q=\pm1$.  The reader interested in getting
to the prescription for spin-4/3 fractional superstring scattering
amplitudes can skip to Section \ref{sIII} which only uses the operator
product expansions summarized at the end of this section.  In a
companion paper \cite{tree2} we will exploit the $\IZ_2$ subgroup of
automorphisms generated by the second transformation in (\ref{autom})
in order to understand the twisted modules of the spin-4/3 algebra.
The next six paragraphs comment on some general features of the
spin-4/3 algebra, after which we begin the detailed analysis of
its untwisted modules.

Since the choice of the algebra (\ref{fsca}) essentially defines the
spin-4/3 fractional superstring, it is worth briefly mentioning the
reasons why one might expect it to give rise both to a sensible and a
computationally manageable string theory.  The representation theory of
the FSC algebra (and related non-local algebras) is well-studied
\cite{ZFpf,ZFft,KMQ,GoS,ALT,AGT,AKT}.  It is known to have a
representation theory similar to that of the conformal and
superconformal algebras.  In particular, it has a series of unitary
minimal representations realized by the coset models $su(2)_4\otimes
su(2)_L/su(2)_{L+4}$ with central charges which accumulate at a
particularly simple $c=2$ model as $L \rightarrow \infty$, analogous to
the free field representation of the conformal algebra with central
charge $1$, or the superconformal algebra with central charge $3/2$.
Presumably a continuum of representations exists for $c\ge2$; some
simple examples will be given later in this paper.

An important feature of the FSC algebra is the appearance of cuts in
the $GG$ OPEs.  Since there is only a single cut on the right-hand side
of each OPE, upon continuation of a correlation function involving,
say, $G^\pm(z)G^\pm(w)$ along a contour interchanging $z$ and $w$ it is
consistent for the correlator to pick up a definite phase.  This
situation is described by saying that the currents $G^\pm$ are {\it
abelianly braided} (or {\it parafermionic}).  Under interchange of $z$
and $w$ (along a prescribed path, say a counterclockwise switch) the
only consistent phase that $G^+$ or $G^-$ can pick up with itself is
$e^{2i\pi/3}$.  The phase that develops upon interchange of $G^+$ with
$G^-$ can be taken to be $e^{-2i\pi/3}$.

Because the operator algebra (\ref{fsca}) is abelianly braided, one can
derive a Ward identity relating correlators with a $G^+G^-$ pair to
ones with the pair removed \cite{ZFpf,ZFft}.  One can then solve for
the structure constants $\lambda^\pm$ by imposing the associativity
condition---independence of which $G^+G^-$ pair we apply the Ward
identity to---on, say, the four-point function $\langle G^{+}(z_{1})
G^{+}(z_{2}) G^{-}(z_{3})G^{-}(z_{4})\rangle$, giving (\ref{lamb}).
The crucial fact that enables us to integrate the Ward identity is the
absence of fractional cuts not allowed by abelian braiding property on
the right-hand side of the OPEs (\ref{fsca}), even among the
``regular'' terms.  This argument is described in more detail in
Appendix C of Ref.~\cite{ALyT}; however, we will not pursue it further
here since we will be able to derive (\ref{lamb}) from the generalized
commutation relations satisfied by the modes of the currents, to be
discussed below.

The abelian braiding of the currents tightly constrains the form of the
FSC algebra.  For example, the appearance of a new primary
dimension-$7/3$ field on the right hand side of the $G^+G^+$ or
$G^-G^-$ OPE would not be consistent with abelian braiding.  On the
other hand, a primary dimension-1 field (and its Virasoro descendents)
could appear in the $G^+G^-$ OPE consistent with abelian braiding as
long as it appeared with opposite sign in the $G^-G^+$ OPE, similar to
the way the spin-1 current enters in the $N=2$ superconformal algebra.
However, we exclude such an operator from (\ref{fsca}) because it can
be shown that such an algebra is only associative for $c=1$, making it
unsuitable for constructing a string theory.

An important consequence of the associativity condition (\ref{lamb}) is
that representations of the FSC algebra can {\it not} be tensored
together to form new representations of the FSC algebra.  Given two
representations of the FSC algebra with the same central charge $c_0$,
and therefore the same structure constants $\lambda^\pm(c_0)$, and
currents $G^\pm_i$, $T_i$ for $i=1,2$, it may seem that one could form
a new representation by tensoring them together.  The tensor-product
algebra would have currents $G^\pm=G^\pm_1+G^\pm_2$, $T=T_1+T_2$,
central charge $c=2c_0$, and structure constant $\lambda^\pm(c_0) =
\lambda^\pm(c/2)$; however, the new central charge and structure
constant are no longer related by (\ref{lamb}), indicating that the
tensor-product representation is not really a representation of the FSC
algebra (\ref{fsca}).  The problem is not that associativity somehow
breaks down for the tensor-product representation, but rather that
taking tensor products introduces new fractional powers among the
regular terms of the OPEs, implying that the braid relations of $G^\pm$
in the tensor-product CFT are different from those in the FSC
algebra.  For example, the first regular term that would appear in the
$G^+G^+$ OPE in the tensor-product CFT is $G^+(z) G^+(w) ~\sim~
(z-w)^0 :\!G^+_1 G^+_2\!:(w)$.  This term and its descendents all
appear with integer powers of $(z-w)$.  Though these terms do not
introduce ``cuts,'' they do nevertheless involve powers of $(z-w)$ that
do not appear (mod integers) among the leading terms of the FSC algebra
OPEs.  The basic lesson is that it is not the OPEs alone that define a
chiral algebra; they must be supplemented by the braid relations
satisfied by the currents.

Thus, the nonlinearity of the FSC algebra, indicated by the dependence
of the structure constants $\lambda^\pm$ on $c$, implies the absence of
tensor-product representations of the algebra.  We will see that it is
this nonlinearity, rather than the fractional dimension of the currents
$G^\pm$, that raises the main obstacles to the existence and
tractability of the spin-4/3 fractional superstring.  Indeed, the
existence of sensible tree scattering amplitudes despite the
nonlinearity of the FSC algebra will appear to occur, in our
formulation, due to an ``accidental'' algebraic cancellation which has
no counterpart in the analogous formulation of bosonic or superstring
scattering amplitudes.

\subsection{The FSC mode algebra}

The physical states of the spin-4/3 fractional superstring are
annihilated by the positive modes of $T$ and $G^\pm$, in analogy to the
``old covariant'' formulation of the bosonic string and superstring in
(super)conformal gauge.  In this section we will define what we mean by
the modes of the $G^\pm$ current, and will derive the algebra that
these modes satisfy.  This discussion will actually only be valid for
certain ``untwisted'' sectors of states analogous to the Neveu-Schwarz
sector of the superstring.  The analysis for the analogs of the Ramond
sector appears in Ref.\ \cite{tree2}.

It will be important in the sequel to know the {\it monodromies} of the
currents $G^\pm$.  The monodromies are the phases picked up when the
insertion point of one current is continued along a closed path around
the insertion point of another current.  We choose this path to be a
simple counterclockwise closed loop, and denote the analytic
continuation of a field $V(z)$ around $W(w)$ by a {\it bypass relation}
\cite{ZFpf}, denoted $V(z) * W(w)$, and illustrated in Fig.~1.  The
monodromies are thus simply the phases acquired upon braiding a pair of
fields twice.  The bypass relations satisfied by the fractional
currents can be read off from the FSC algebra OPEs (\ref{fsca}):
	\begin{eqnarray}
   \label{currbypass} T(z) * T(w) &=& T(z)\, T(w), \nonumber\\ T(z) *
	G^\pm(w) &=& T(z)\, G^\pm(w), \nonumber\\ G^\pm(z) * G^\pm(w)
	&=& {\rm e}^{-2i\pi/3} G^\pm(z)\, G^\pm(w), \nonumber\\
	G^\pm(z) * G^\mp(w) &=& {\rm e}^{+2i\pi/3} G^\pm(z)\,
	G^\mp(w).
	\end{eqnarray}

As was pointed out in Ref.\ \cite{ZFft}, the FSC algebra (\ref{fsca})
has a $\IZ_3$ symmetry that is useful in organizing its
representation theory.  In particular, the currents $G^{+}$ and $G^{-}$
can be assigned $\IZ_3$ charges $q=1$ and $-1$, respectively, while
the energy-momentum tensor $T$ (as well as the identity) has charge
$q=0$.  It is natural to assume that, since the FSC algebra is supposed
to be an organizing symmetry of our theory, all the fields in a
representation will have definite $\IZ_3$ charges, and that these
fields will have the same monodromies with the FSC currents as the
currents have with themselves.  These conditions define the class of
untwisted representations of the FSC algeba.  Along with the
parafermionic nature of the FSC algebra, these properties enable one to
learn much about the structure of these representations \cite{ZFpf}.

So, assume the state space of FSC algebra representations falls into
sectors ${\cal U}_q$ labelled by their $\IZ_3$ charge.  The
currents $G^{+}$ and $G^{-}$ have $\IZ_3$ charges $q=+1$ and
$q=-1$, respectively, and so act on the Fock space sectors as $G^\pm:
{\cal U}_q \rightarrow {\cal U}_{q\pm 1}$.  A state $\chi_q\in{\cal
U}_q$ obeys the bypass relations
	\begin{eqnarray}
   \label{bypass} T * \chi_q &=& T\, \chi_q, \nonumber\\ G^\pm *
	\chi_q &=& {\rm e}^{\mp2i\pi q/3} G^\pm\, \chi_q.
	\end{eqnarray}
Note
that these monodromies are consistent with (\ref{currbypass}) and the
$\IZ_3$ charge assignments of $T$ and $G^\pm$.  These bypass
relations imply that we can define the mode expansions of $G^{+}$ and
$G^{-}$ acting on any state $\chi_q$ by
  \begin{eqnarray}
   \label{Gmode} G^{+}(z)\chi_q(0)&=&\sum_{n\in\IZ} z^{n-q/3}
   G^{+}_{-1-n-(1-q)/3}\chi_q(0)~,\nonumber\\
  G^{-}(z)\chi_q(0)&=&\sum_{n\in\IZ} z^{n+q/3}
   G^{-}_{-1-n-(1+q)/3}\chi_q(0)~.
	\end{eqnarray}
The point is simply that the
powers appearing on the right-hand side are the only ones which pick up
phases consistent with (\ref{bypass}) upon a counterclockwise
continuation of $z$ around $0$.  The moding of the currents labels the
(operator) coefficients of these terms with the convention that the
value of the mode number is the negative of the dimension of the mode
operator.  The mode expansions (\ref{Gmode}) can be inverted to give
	\begin{eqnarray}
   \label{edomG}
	  G^{+}_{n-(1-q)/3}\chi_q(0)&=&\oint_\gamma{{\rm d}z
	   \over2\pi i} z^{n+q/3}G^{+}(z)\chi_q(0)~,\nonumber\\
	  G^{-}_{n-(1+q)/3}\chi_q(0)&=&\oint_\gamma{{\rm d}z
	   \over2\pi i} z^{n-q/3}G^{-}(z)\chi_q(0)~.
	\end{eqnarray}
Here,
$\gamma$ is a contour encircling the origin once, where $\chi_q(0)$ is
inserted.  The allowed modings of the currents in the different $\IZ_3$
sectors are summarized in Fig.~2.  Note that the action of a
$G^\pm$ mode on a state in a given sector will map it to a different
sector, where, in general, different modings are allowed.

Following the arguments of Ref.~\cite{ZFpf}, the generalized
commutation relations (GCR) satisfied by the current modes of the FSC
algebra (\ref{fsca}) can be derived.  We briefly review this argument
by deriving the GCRs of the modes of $G^+$ with $G^-$.  The general
procedure for deriving GCRs for the modes of any abelianly braided
operators should be clear from this example.

Consider the integral
\begin{equation}
	\label{II}
 \oint_{\gamma}{{\rm d}z\over2\pi i}
  \oint_{\delta}{{\rm d}w\over2\pi i} z^{m+q/3}
  w^{n-q/3}(z-w)^{p+2/3} G^+(z)G^-(w)\chi_q(0)~,
\end{equation}
where $m$, $n$ and $p$ are arbitrary integers.  The contours $\gamma$
and $\delta$ encircle the origin, with $\delta$ inside $\gamma$.  The
fractional parts of the exponents in the integrand are chosen so that
the whole integrand is single-valued in both the $z$- and $w$-planes.
This is possible only because of the abelian nature of the $G^+G^-$
OPE.  Evaluate this integral by letting $\delta$ shrink down to a small
circle near to the origin.  In this limit, expand the $(z-w)^\alpha$
factor as $(z-w)^\alpha =\sum_{\ell=0}^\infty C^{(\alpha)}_\ell
z^{\alpha-\ell}w^\ell$, where $C^{(\alpha)}_\ell$ are the appropriate
fractional binomial coefficients:
\begin{equation}
	\label{cexp}
 C^{(\alpha)}_\ell=(-1)^\ell{\alpha\choose\ell}.
\end{equation}
Inserting this expansion into (\ref{II}) and using the mode definitions
(\ref{edomG}) gives $\sum_{\ell=0}^\infty C^{(p+2/3)}_\ell
G^+_{m+p-\ell+(1+q)/3} G^-_{n+\ell-(1+q)/3} \chi_q(0)$.  The integral
(\ref{II}) can also be evaluated in another way, by first deforming the
$\gamma$ contour so that it lies inside $\delta$.  Upon performing this
deformation, one picks up in the usual way two contributions
corresponding to the same integral with $\gamma$ and $\delta$
interchanged, and a contribution where the $\gamma$ contour encircles
the $G^-$ insertion at the point $w$ on the $z$-plane; see Fig.~3.  The
contribution with the $\gamma$ and $\delta$ contours interchanged is
evaluated in the same way as outlined above after interchanging
$G^+(z)$ and $G^-(w)$ as well as $z$ and $w$ in the $(z-w)^{p+2/3}$
factor.  Taking care to perform these interchanges along equivalent
paths in the complex plane gives an overall phase ${\rm
e}^{i\pi(-2/3)}\times{\rm e}^{i\pi(p+2/3)} = (-1)^p$ (where the abelian
braiding of $G^+$ with $G^-$ has been used).  The second contribution,
where $\gamma$ only encircles the point $w$ in the $z$-plane, is
evaluated by letting this contour shrink to a small circle around $w$
and replacing $G^+(z) G^-(w)$ by their OPE.  The value of the integer
$p$ in the integrand controls the number of terms in the OPE that
contribute.  For example, taking $p=-1$ and assembling the three
contributions shown in Fig.~3 results in the generalized commutation
relation for the $G^+$ and $G^-$ modes given below in Eq.~(\ref{gcr}).

Alternatively we could have chosen another value of $p$, which would
pick up different contributions from the $G^+G^-$ OPE.  It is clear
that by letting $p$ take more negative values, more complicated GCRs
involving more terms from the $G^+G^-$ OPE can be obtained.  By
conformal invariance, this tower of GCRs is consistent.  Indeed, the
GCR obtained with $p=p_0$ can be derived from the GCR with $p=p_0-1$
using the binomial coefficient identity $
C^{(\alpha)}_\ell-C^{(\alpha)}_{\ell-1}~=~C^{(\alpha+1)}_\ell~.$ So,
there are many GCRs that can be derived from a single OPE, depending on
how many terms on the right hand side of the OPE one wishes to
include.  We will include only the singular terms, shown in
eq.~(\ref{fsca}).

With this choice, the FSC algebra GCRs become \cite{ZFft}
\begin{eqnarray}
   \label{gcr}
 \sum_{\ell=0}^\infty C^{(-2/3)}_\ell\left[
  G^{+}_{{q\over3}+n-\ell}G^{+}_{{2+q\over3}+m+\ell}-
  G^{+}_{{q\over3}+m-\ell}G^{+}_{{2+q\over3}+n+\ell}\right]
  &=& {\lambda^+\over2}(n-m)G^{-}_{{2+2q\over3}+n+m},\nonumber\\
 \sum_{\ell=0}^\infty C^{(-2/3)}_\ell\left[
  G^{-}_{-{q\over3}+n-\ell}G^{-}_{{2-q\over3}+m+\ell}-
  G^{-}_{-{q\over3}+m-\ell}G^{-}_{{2-q\over3}+n+\ell}\right]
  &=& {\lambda^-\over2}(n-m)G^{+}_{{2-2q\over3}+n+m},\nonumber\\
 \sum_{\ell=0}^\infty C^{(-1/3)}_\ell\left[
  G^{+}_{{1+q\over3}+n-\ell}G^{-}_{-{1+q\over3}+m+\ell}+
  G^{-}_{-{2+q\over3}+m-\ell}G^{+}_{{2+q\over3}+n+\ell}\right]
 &=& L_{n+m} \nonumber\\
  \mbox{}+{3c\over16} \left(n+1+{q\over3}\right)
  \!\!\!\!\!\!\!&&\!\!\!\!\!\!
  \left(n+{q\over3}\right)\delta_{n+m},
\end{eqnarray}
where these expressions are understood to be acting on a state in
${\cal U}_q$.  Because of the infinite sum on the left-hand sides, the
mode algebra in Eq.~(\ref{gcr}) is not a graded Lie algebra, but a new
algebraic structure on the string world-sheet.  This infinite sum is a
reflection of the fractional dimension of the current $G$ and the
resulting cuts in its OPEs.  Though the GCRs look complicated, they are
as useful as the familiar (anti)commutators of the (super)Virasoro
algebra.  The reason for this is that the integer $\ell$ appearing in
the infinite sum is bounded from below. Acting on any state of fixed
conformal dimension, the left-hand sides of the GCRs will have only a
finite number of non-zero terms since for large enough $\ell$ the
$G^\pm$ modes will annihilate the state.  Examples of the use of the
GCRs will be given later in this section.

For completeness, we also write down the standard commutators following
from the conformal algebra and the fact that $G^\pm$ are
dimension-$4/3$ Virasoro primary fields:
	\begin{eqnarray}
   \label{vircom}
	 \left[L_m,L_n\right]&=&(m-n)L_{m+n}
	 +{c\over12}(m^3-m)\delta_{m+n}~,\nonumber\\
	 \left[L_m,G^\pm_r\right]&=&
	 \left({m\over3}-r\right)G^\pm_{m+r}~,
	\end{eqnarray}
where the moding $r$
is the one appropriate to whichever $\IZ_3$ sector the $G^\pm$
currents are acting on, and the $L_n$ are the standard modes of the
stress-energy tensor defined by
	\begin{equation}
	\label{Tmode}
	 T(z)\chi_q(0)=\sum_{n\in\IZ} z^{-n-2} L_n\chi_q(0),
	\end{equation}
independent of $q$.

\subsection{FSC highest-weight modules}

A {\it highest-weight state} (or {\it primary state}) $\ket{\chi}$ of
the FSC algebra is a state which is annihilated by all the positive
modes of $T$ and $G^\pm$:
	\begin{equation}
	\label{hws} L_n\ket{\chi}=G^\pm_{n/3}\ket{\chi}=0, \qquad n\in
	\IZ>0.
	\end{equation}
 A {\it highest-weight module} of the FSC algebra is
a highest-weight state $\ket{\chi}$ along with all its {\it
descendent} states formed from $\ket{\chi}$ by the action of creation
(or zero) modes of $T$ and $G^\pm$.  It is easy to see from the
$LG^\pm$ commutator (\ref{vircom}) that any sequence of $L$ and $G^\pm$
creation modes can be reordered (to a different set) so that the $L$'s
are to the right of the $G^\pm$'s.  Thus the general descendent of
$\ket{\chi}$ can be written
	\begin{equation}
	\label{hwm} G^\pm_{r_1}\cdots G^\pm_{r_p} L_{n_1}\cdots
	L_{n_q}\ket{\chi}, \quad r_i\in\IZ/3\le0, n_j\in
	\IZ\le0.
	\end{equation}
 States annihilated by the positive $L_n$ modes are
{\it Virasoro primaries}, while states created from a primary state by
the action of the $L_{-n}$ modes alone are {\it Virasoro descendents}.
In general, we will use the term ``descendent'' without modifier to
mean descendents with respect to the FSC algebra.  Thus, descendent
states can be Virasoro primary or Virasoro descendent.

Before describing the properties of the FSC modules in detail, let us
first outline how the FSC mode algebra (\ref{gcr}) is used in
practice.  The basic problem that the mode algebra should answer is how
any sequence of $G^\pm$ and $L$ creation or annihilation modes in which
the sum of all the modings is non-positive, can be written as a
descendent state as in (\ref{hwm}).  For the $L$'s alone, this follows
from the Virasoro algebra (\ref{vircom}) by repeatedly commuting the
positively-moded $L$'s to the right until they annihilate the
highest-weight state $\chi$.  The analogous operation for the $G^\pm$
modes is less clear due to the infinite sums in their GCRs
(\ref{gcr}).

To explain how this can be done, we first need to show a basic property
of the GCR algebra (\ref{gcr}).  For any highest-weight state $\chi$,
and any $r_i\in\IZ/3$,
	\begin{equation}
	\label{lemma} G^\pm_{r_1}\cdots G^\pm_{r_p}\ket{\chi}=0 \qquad
	{\rm if}\ \sum_{i=1}^p r_i >0.
	\end{equation}
 Consider the $p=2$ case
first, where we want to show that $ G^\alpha_r G^\beta_s\ket{\chi}=0 $
if $r+s>0$, where $\alpha$ and $\beta$ are $\pm$. If $s>0$ the
expression vanishes because $\chi$ is highest-weight, so we only need
to examine the case $r>-s\ge0$.  {}From the form of the GCRs
(\ref{gcr}) it follows that
	\begin{equation}
	\label{genericgcr} \sum_{\ell=0}^\infty C^{(a)}_\ell\left[
	G^\alpha_{r-\ell} G^\beta_{s+\ell} \pm
	G^\beta_{s'-\ell}G^\alpha_{r'-\ell}\right]\ket{\chi} \sim
	\left(G^\gamma_{r+s}\ {\rm or}\ L_{r+s}\right)\ket{\chi},
	\end{equation}
where $s'=s-1/3$ or $s-2/3$ and $r'=r+1/3$ or $r+2/3$.  Since $\chi$ is
highest-weight, and using $r>0$, $r+s>0$ and $C^{(a)}_0=1$, the above
expression reduces to the finite sum
	\begin{equation}
	 G^\alpha_rG^\beta_s\ket{\chi}=\sum_{\ell=1}^{-s}
	C^{(a)}_\ell G^\alpha_{r-\ell}G^\beta_{s+\ell} \ket{\chi}.
	\end{equation}
Note that all the terms in the sum on the right-hand side are of the
same form as the original term on the left-hand side, except that the
modings of $G^\beta$ are less negative.  Repeatedly applying the same
argument to these terms, one can eventually show that they are zero.
For the general case, $p>2$, perform the same argument on
$G^\alpha_{r_1}G^\beta_{r_2}$ in (\ref{lemma}) using the $p=2$ result
and proceed by induction on $p$.

This shows how to perform the operation analogous to ``commuting a mode
to the right'' using the GCRs (\ref{gcr}).  Namely, the GCRs relate the
product of the mode in question and its neighbor to the right to an
infinte sum of products of modes as in (\ref{genericgcr}).  However, by
(\ref{lemma}) all but a finite number of these terms vanish, and
repeated applications of the GCRs on the remaining terms will
eventually convert them all to creation operators.

As an example of the use of the GCRs, we derive the associativity
constraint (\ref{lamb}) on the structure constants $\lambda^\pm$.
Consider a highest-weight state $\chi$ with $\IZ_3$-charge $q=0$
satisfying $L_0\ket{\chi}=h\ket{\chi}$ and its descendent state
	\begin{equation}
	\label{chiprime} \ket{\chi'}=G^-_0G^+_0G^+_{-1/3}\ket{\chi}.
	\end{equation}
We can simplify $\chi'$ by using the $G^+G^+$ GCR of Eq.~(\ref{gcr})
with $q=0$, $m=-1$, $n=0$ to find (since $\chi$ is highest-weight)
	\begin{equation}
	 \ket{\chi'}={\lambda^+\over2}G^-_0G^-_{-1/3}\ket{\chi}.
	\end{equation}
 Now using the $G^-G^-$ GCR with the same values of $q$,
$m$, and $n$ gives
	\begin{equation}
	\label{chiassoci} \ket{\chi'} = {\lambda^+\lambda^-\over4}
	G^+_{-1/3} \ket{\chi}.
	\end{equation}
 Alternatively, we can try to
simplify (\ref{chiprime}) with the $G^-G^+$ GCR.  Acting on the state
$G^+_{-1/3}\ket{\chi}$ (which is in the $q=+1$ sector) with $q=1$,
$m=1$, $n=-1$, this GCR gives
	\begin{eqnarray}
   \label{chiassoc} \left( G^+_{-1/3} G^-_{1/3} + G^-_0 G^+_0 \right)
	G^+_{-1/3} \ket{\chi} &=& \left( L_0 - {c\over24} \right)
	G^+_{-1/3} \ket{\chi} \nonumber\\ &=& \left (h + {1\over3} -
	{c\over24} \right) G^+_{-1/3} \ket{\chi},
	\end{eqnarray}
where we have
used (\ref{lemma}) to remove all but two terms from the infinite sum.
Using the $G^-G^+$ GCR again with $q=0$, $m=1$, $n=-1$ on $\chi$ shows
$G^-_{1/3}G^+_{-1/3}\ket{\chi} = L_0\ket{\chi} = h\ket{\chi}$, which,
when substituted in (\ref{chiassoc}), shows that
	\begin{equation}
	\label{chiassocii} \ket{\chi'} = \left( {1\over3} - {c\over24}
	\right) G^+_{-1/3} \ket{\chi}.
	\end{equation}
Comparing (\ref{chiassoci}) and (\ref{chiassocii}) determines $\lambda^+
\lambda^-$.

\subsubsection{S-modules}

The basic properties of the FSC modules built on highest-weight states
with $\IZ_3$ charge $q=0$ follow from the non-vanishing modings of
$G^\pm$ and their GCRs.  The modules based on these states are called
``S-modules'' \cite{ZFft}.  On states of $\IZ_3$ charge zero, the
only non-vanishing modings $G^\pm_r$ have $r\in\IZ-1/3$.  In
particular, there is no (non-vanishing) $G^\pm$ zero mode.  The only
zero mode is thus $L_0$, and let us choose highest-weight states $W_s$
for these modules to be $L_0$ eigenstates of eigenvalue $h_s$.  A given
S-module will be completely determined by $h_s$ and $c$, the value of
the central charge in whatever representation of the FSC algebra we are
considering.

The first descendents of $\ket{W_s}$ are $G^\pm_{-1/3}\ket{W_s}$, which
are Virasoro primary states of dimension $h_s+{1\over3}$ with $\IZ_3$
charges $q=\pm1$.  To create further descendents from these states by
the action of $G^\pm_r$ modes requires either $r\in\IZ$ or
$r\in\IZ-{2\over3}$ (see Fig.~2).  Proceeding in this way, it is
easy to see that the S-module will consist of $q=0$ states of dimension
$h_s+n$, and $q=\pm1$ states of dimension $h_s+{1\over3}+n$, where
$n\ge0$ is an integer.

One may try to build an infinite series of S-module descendent states
of given conformal dimension by the action of the $G^\pm_0$ modes.  For
example, at dimension $h_s+{1\over3}$ we have $(G^-_0G^+_0)^p
G^+_{-1/3} \ket{W_s}$, for any non-negative integer $p$.  However, the
GCRs (\ref{gcr}) show that these states are not independent: they are
equal to $(\lambda^+\lambda^-/4)^pG^+_{-1/3}\ket{W_s}$.  In general,
the problem of finding a basis of independent states at each level can
be complicated.  The number of independent states at each level has
been determined in Ref.~\cite{AKT}.

The structure of the S-module descendents can be
summarized by the operator product expansions of the FSC current
$G^\pm$ with the $W_s$ primary state and its descendents.  They are
	\begin{eqnarray}
   \label{Smod} G^\pm(z)W_s&=&{V^\pm_s\over z}+\ldots\nonumber\\
	G^\pm(z)V^\pm_s&=& \left({\lambda^\pm\over2}\right) {1\over
	z^{4/3}}
	 \left\{V^\mp_s + {2z\over 3h_s+1} \partial V^\mp_s \right\}+
	 {\widetilde V^\mp_s\over z^{1/3}} +\ldots \nonumber\\
	G^\pm(z)V^\mp_s&=& {h_s\over z^{5/3}} \left\{W_s + {z\over2h_s}
	 \partial W_s\right\}\pm{\widetilde W_s\over z^{2/3}}+\ldots
	\\ G^\pm(z)\widetilde W_s &=&
	\pm\left(2-8h_s-c\over6\right) {1\over z^2} \left\{ V^\pm_s +
	{z\over 3h_s+1} \partial V^\pm_s \right\} \pm \lambda^\pm
	{\widetilde V^\pm_s \over z} + {h_s\over 3}
	{{\widetilde{\widetilde V}}^\pm_s \over z} + \ldots \nonumber
	\end{eqnarray}
For ease of writing, we have inserted the S-module
vertex operators $W_s$ {\it etc.}\ at the origin of the complex plane
and have dropped their arguments.  $V^\pm_s$, $\widetilde V^\pm_s$, and
${\widetilde{\widetilde V}}^\pm_s$ are new Virasoro (though not FSC)
primaries of conformal dimension $h_s+{1\over3}$, $h_s+{4\over3}$, and
$h_s+{4\over3}$, respectively, while $\widetilde W_s$ is a dimension
$h_s+1$ Virasoro primary.  They are defined by
	\begin{eqnarray}
   \label{Sdesc} \ket{V^\pm_s} &=& G^\pm_{-1/3}\ket{W_s},\nonumber\\
	\ket{\widetilde V^\pm_s} &=& G^\mp_{-1}\ket{V^\mp_s}
	 -{\lambda^\mp\over3h_s+1}L_{-1}\ket{V^\pm_s},\nonumber\\
	\ket{{\widetilde{\widetilde V}}^\pm_s} &=&
	2G^\pm_{-4/3}\ket{W_s} - {5\over 3h_s+1} L_{-1}\ket{V^\pm_s} ,
	\nonumber\\ \ket{\widetilde W_s} &=& \left(G^+_{-2/3}G^-_{-1/3}
	- G^-_{-2/3}G^+_{-1/3}\right)\ket{W_s}.
	\end{eqnarray}
The various
coefficients appearing in (\ref{Smod}) were determined by the FSC mode
algebra (\ref{gcr}).  For example,
	\begin{eqnarray}
	 G^+_{1/3}\ket{V^-_s} &=& G^+_{1/3} G^-_{-1/3} \ket{W_s}
	\nonumber\\ &=& L_0\ket{W_s} = h_s \ket{W_s},
	\end{eqnarray}
giving the
first coefficient in the $G^+V^-_s$ OPE in (\ref{Smod}).  Here the
$G^+G^-$ GCR (\ref{gcr}) was used in the second equality.

We should think of $V^\pm_s$ as forming a ``fractional supermultiplet''
with $W_s$. Note that $W_s$ is single-valued with respect to the
currents $T$ and $G^\pm$, while $V^\pm_s$ have cuts with the fractional
current, reflecting the ``fractional statistics'' of $V^\pm_s$ on the
world-sheet.  Summarizing, S-module are
characterized by the fields $(W_s,V^\pm_s)$ belonging to a fractional
superconformal multiplet with conformal dimensions
$(h_s,h_s+{1\over3})$ and with world-sheet statistics ({\it bosonic~,
fractional\/}).

\subsubsection{D-modules}

FSC modules built on highest-weight states with $\IZ_3$ charge $q=\pm1$
are called ``D-modules'' \cite{ZFft}.  The non-vanishing modings
$G^\pm_r$ on $q=\pm1$ states have $r\in\IZ$ or $r\in\IZ-2/3$.
If we choose the highest-weight states of the D-module to have
conformal dimension ($L_0$ eigenvalue) $h_d$, it follows that
descendents will have dimension $h_d+n$ in the $q=\pm1$ sectors and
dimension $h_d+{2\over3}+n$ in the $q=0$ sector, for $n$ a non-negative
integer.  However, the structure of D-modules is more complicated than
that of S-modules because of the action of the $G^\pm_0$ modes on the
highest-weight state.

Consider a highest-weight state $W^+$ with $q=+1$, and conformal
dimension $h_d$.  In general, this state will be degenerate with
another state $\ket{W^-}=G^+_0\ket{W^+}$ which also has dimension
$h_d$, but has charge $q=-1$.  The $G^+G^-$ GCR in (\ref{gcr}) implies
that $G^-_0\ket{W^-}=(h_d-{c\over24})\ket{W^+}$, so fractional
highest-weight states are doubly degenerate.  It is convenient to
normalize these states to satisfy
	\begin{equation}
	\label{Dint} G^\pm_0\ket{W^\pm_d}=\Lambda^\pm \ket{W^\mp_d},
	\end{equation}
where
	\begin{eqnarray}
   \label{Dlamb} \Lambda^+& =& \Lambda^- =
	\sqrt{h_d-{c\over24}},\qquad{\rm for}\quad c<24h_d , \nonumber\\
	\Lambda^+& =& -\Lambda^- = \sqrt{{c\over24}-h_d},\qquad{\rm
	for}\quad c>24h_d .
	\end{eqnarray}
The main properties of D-modules are summarized by the OPEs of the
currents $G^\pm$ with the highest weight vertex operators $W^\pm_d(z)$
of dimension $h_d$ and their first descendent operators
$V^{(\pm)}_d(z)$ of dimension $h_d+{2\over3}$:
	\begin{eqnarray}
   \label{Dmod} G^\pm(z)W^\pm_d&=&{\Lambda^\pm\over
	z^{4/3}}\left\{W^\mp_d + {2z\over3h_d} \partial W^\mp_d\right\}
	+{{\widetilde W^\mp_d}\over z^{1/3}}+\ldots\nonumber\\
	G^\mp(z) W^\pm_d&=&{1/2\over z^{2/3}}\left\{ V^{(+)}_d \mp
	V^{(-)}_d \right\} +\ldots\nonumber\\
	G^\pm(z)V^{(+)}_d&=&\left(h_d + {c\over12} +
	\lambda^\pm\Lambda^\mp\right) {1\over z^2} \left\{W^\pm_d +
	{z\over3h_d} \partial W^\pm_d\right\}\nonumber\\
	&&\qquad\mbox{}-\left(\Lambda^\pm -{1\over2}
	\lambda^\pm\right) {\widetilde W^\pm_d\over z} +\ldots
	\nonumber\\ G^\pm(z)V^{(-)}_d&=&\mp\left(h_d +
	{c\over12} - \lambda^\pm\Lambda^\mp\right) {1\over z^2}
	\left\{W^\pm_d + {z\over3h_d} \partial
	W^\pm_d\right\}\nonumber\\ &&\qquad\mbox{} \pm
	\left(\Lambda^\pm + {1\over2}\lambda^\pm\right) {\widetilde
	W^\pm_d\over z} +\ldots
	\end{eqnarray}
The first OPE defines the two
(Virasoro primary) descendent operators of conformal dimension $h_d+1$
and $\IZ_3$ charge $q=\pm1$
	\begin{equation}
	\label{Ddesc} \ket{\widetilde W^\pm_d} = G^\mp_{-1}\ket{W^\mp_d} -
	{2\Lambda^\mp \over3h_d} L_{-1}\ket{W^\pm_d},
	\end{equation}
 while the
second OPE defines the $\IZ_3$ charge $q=0$ Virasoro primary
descendents of conformal dimension $h_d+{2\over3}$
	\begin{equation}
	\label{Ddesc2} \ket{V^{(\pm)}_d} = G^+_{-2/3}\ket{W^-_d}
	\pm G^-_{-2/3}\ket{W^+_d} .
	\end{equation}
 We have put the $\pm$
superscript on the $V_f$ descendent states in parenthesese to emphasize
that they do {\it not} refer to the $\IZ_3$ charge of these states.  We
have chosen the particular definition (\ref{Ddesc2}) of $V^{(\pm)}_d$
for later convenience.  Just as in the S-module case, the coefficients
of the OPEs (\ref{Dmod}) are determined from the FSC mode algebra
(\ref{gcr}).

We think of $V^{(\pm)}_d$ as forming a fractional supermultiplet with
$W^\pm_d$. The fractional currents $G^\pm$ are single-valued with
respect to the $q=0$ descendent $V^{(\pm)}_d$ but has cuts with the
$q=\pm1$ highest-weight states $W^\pm_d$.  In summary, D-modules are
characterized by the central charge $c$ and the conformal dimension
$h_d$ of their two highest-weight fields.  The D-module fractional
supermultiplets are always of the form of a set of fields
$(W^\pm_d,V^{(\pm)}_d)$ with conformal dimensions $(h_d,h_d+{2\over3})$
and with world-sheet statistics ({\it fractional~, bosonic}).

\section{\label{sIII}Tree scattering amplitudes}

In this section we formulate tree-level scattering amplitudes of
physical states in the spin-4/3 fractional superstring, and show that
these amplitudes obey spurious state decoupling and duality
properties.  In what follows, we construct open string scattering
amplitudes.  Closed string scattering amplitudes at tree level are
easily formed by combining two open string amplitudes using a
level-matching condition for left- and right-movers \cite{KLT}.  The
construction we use is closely analogous to that of open
Ramond-Neveu-Schwarz superstring tree amplitudes in the ``old
covariant'' formalism \cite{NST}.

We take the physical states of the fractional superstring to be
highest-weight states of the FSC algebra.  Thus the physical state
conditions are the requirement that the positive (annihilation) modes
of the FSC currents vanish when acting on physical states.  This
definition of physical states can be motivated as follows.  In the
usual superstring, the physical state conditions are constraints
following from gauge-fixing the local world-sheet symmetry.
Classically these constraints in the superconformal gauge are given by
the vanishing of the energy-momentum tensor and superconformal
current.  The full local world-sheet symmetry of the spin-4/3
fractional superstring is unknown, though it should include invariances
under reparametrizations and Weyl rescalings of the world-sheet.
Assume that some analog of the superconformal gauge exists in the
fractional superstring, giving rise to an algebra of constraints
generated by the vanishing of $T(z)$ and the fractional superconformal
currents $G^\pm(z)$.  In other words, assume that the fractional
superconformal algebra is the quantum version of some classical
constraint algebra.  Thus, although we do not know of any classical
local symmetry on the world-sheet that gives rise to a spin-4/3 current
as a constraint upon gauge-fixing, we nevertheless assume the weak
physical state conditions
	\begin{equation}
	\label{qcon} \langle\psi\vert T(z) \ket{\phi}=
	\langle\psi\vert G^\pm(z) \ket{\phi}=0~,
	\end{equation}
 for any
physical states $\ket{\phi}$ and $\ket{\psi}$.
Just as the stress-energy constraint is satisfied if the physical
states are defined to be those states annihilated by the stress-energy
modes $L_n$ with $n>0$, we can factorize the fractional current
constraint
by demanding that all physical states are annihilated by non-negative
modes of $G^\pm$.  Thus, a physical state $\ket{\phi}$ should
satisfy
	\begin{eqnarray}
   \label{psc} (L_n-h\delta_{n,0})\ket{\phi}&=&0~,\quad 0\le n\in
	\IZ\ ,\nonumber\\ G^\pm_r\ket{\phi}&=&0~,\quad 0 < r\in\IZ/3\ ,
	\end{eqnarray}
where $r$ is the appropriate moding depending on the $\IZ_3$ charge
of the state, as in Eq.\ (\ref{Gmode}).
Note that, by (\ref{vircom}), all
the positively-moded constraints can be generated from those of the set
$\{L_1, L_2, G^\pm_{1/3}, G^\pm_{2/3}, G^\pm_1, G^\pm_{4/3}\}$.

{}From the physical state conditions (\ref{psc}) it is clear that
physical states are highest-weight states of the FSC algebra.  If the
state has $\IZ_3$ charge $q=0$ it is the highest weight state of an
S-module with conformal dimension $h_s=h$.  If the state has $\IZ_3$
charge $q=\pm1$ it is the highest-weight state of a D-module with
$h_d=h$.  Here $h$ is the ``intercept'',  a normal ordering constant in
the definition of $T$.  The value of this intercept should be
determined by demanding consistency (unitarity, anomaly cancellation)
of the string scattering amplitudes.

The above argument suggests that there should also be a $G^\pm_0$
physical state condition for D-module states (integer moding is not
allowed on S-module highest-weight states---see Fig.\ 2).  However,
since the FSC algebra essentially determines the action of the
$G^\pm_0$ modes on the two highest-weight states of a D-module in terms
of their common $L_0$ intercept $h_d$ as described in Section 2.2.2, we
find that we do not need to impose any extra zero-mode physical state
condition to those of Eq.\ (\ref{psc}).

The standard properties of spurious and null states follow from the
physical state conditions.  A state $\ket{s}$ obeying the zero-mode
conditions in Eq.~(\ref{psc}) is called a spurious state if it is
orthogonal to all physical states.  Such a state can be written as
	\begin{equation}
	\label{spur} \langle s| = \sum_{n>0} \langle \chi_n| L_n +
	\sum_{r>0} \langle \psi^\pm_r| G^\pm_r\ ,
	\end{equation}
 in terms of
some other states $\ket{\chi_n}$ and $\ket{\psi^\pm_r}$.  Since
$\ket{s}$ is orthogonal to all physical states, the operator
$\ket{s}\langle s|$ must annihilate all physical states.  Since the
physical state conditions (\ref{psc}) are the only restriction on a
generic physical state, it follows that $\ket{s}\langle
s|=\sum_{n>0}X_nL_n + \sum_{r>0}\Psi^\pm_r G^\pm_r$ for some operators
$X_n$ and $\Psi^\pm_r$.  Eq.~(\ref{spur}) follows with $\langle \chi_n|
= \langle s| X_n$ and $\langle \psi^\pm_r| = \langle s| \Psi^\pm_r$.
All states not satisfying the physical state conditions must have a
spurious component.  A physical state can itself be spurious, in which
case it is a null state (since it is orthogonal to itself), and should
decouple from all scattering amplitudes.  Thus, the decoupling of all
spurious states from scattering amplitudes of physical states is a
prerequisite for a sensible interpretation of those amplitudes.

We formulate scattering amplitudes of physical states by first
satisfying the requirement of conformal invariance on the string
world-sheet.  This essentially ensures decoupling of spurious states
which are created solely by modes of $T$ in (\ref{spur}).  This
consideration and the resulting description of scattering amplitudes is
identical to that encountered in the ``old covariant'' formulation of
bosonic string amplitudes \cite{NST}.  We briefly describe heuristic
arguments that lead to a prescription for fractional superstring
scattering; however, this presription is only really justified by the
spurious state decoupling argument which then follows.

The world-sheet in an open string tree scattering process is
conformally equivalent to a unit disc with vertex operators $V(x)$
representing the asymptotic scattering states inserted at points on the
boundary.  Since we must be able to integrate these vertex operators
over their insertion positions, they must be dimension-one operators in
the two-dimensional world-sheet theory.  Furthermore, as in the bosonic
string, they must be Virasoro primary operators.  We can conformally
map the disk to the complex upper half-plane, fixing the positions of
three of the vertex insertions at $\infty$, 1, and 0 on the real axis,
with the remaining insertions at points $1<x_i<\infty$.  The boundary
conditions at the ends of the string (the real axis) can be implemented
by the standard trick of extending the amplitude to the full complex
plane so that holomorphic functions on the upper half-plane correspond
to left-moving excitations of the open string and holomorphic functions
on the lower half-plane correspond to right-moving modes.  The boundary
conditions then imply the continuity of these functions across the real
axis.  This picture is suitable for writing the string amplitude as a
correlator of holomorphic operators only with a radial-ordering
prescription:
	\begin{equation}
	\label{Aone} {\cal A}_N= \int {dx_3\cdots dx_{N-1}\over x_3\cdots
	x_{N-1}} \langle  V_N| V_{N-1}(x_{N-1})\cdots V_2(1)\ket{
	V_1}\ ,
	\end{equation}
 where the ``in'' and ``out'' states are the
insertions at $x_1=0$ and $x_N=\infty$, and the integration is over all
$x_i$ preserving the order $1<x_3<\cdots<x_{N-1}<\infty$.  A vertex
insertion at $x$ can be rewritten as $ V(x) = x^{L_0} V(1) x^{-L_0}$,
and the positions of the insertions explicitly integrated over to give
the amplitude in the form
	\begin{equation}
	\label{Aprop} {\cal A}_N= \langle V_N| V_{N-1}(1)
	\widetilde\Delta\ldots\widetilde\Delta
	 V_2(1)\ket{ V_1}
	\end{equation}
 where the propagator is
$\widetilde\Delta=(L_0-1)^{-1}$.

{}From the presentation on the disk, it is clear that ${\cal A}_N$
should be invariant under cyclic permutations of the vertex ordering.
The cyclic symmetry of open string amplitudes is known as ``duality''.
It can be formulated in the picture corresponding to (\ref{Aone}) as
the requirement that after passing the $V_N$ vertex to the right
through all the other vertices the value of ${\cal A}_N$ must be
unchanged.  Now, suppose the string describes particles in some flat
space-time with coordinate fields $X^\mu(z)$.  Then, by translation
invariance, the general vertex in (\ref{Aone}) will be of the form
	\begin{equation}
	 V_i(x)=V_0(k_i,x){\rm e}^{ik_i\cdot X(x)},
	\end{equation}
 where $V_0$
depends only on derivatives of $X$ (as well as any other conformal
fields on the world-sheet).  Upon commuting the $e^{ikX}$ factors of
two vertices, one picks up the phase ${\rm exp}[i\pi k_i\cdot
k_j\epsilon(x_i-x_j)]$, where $\epsilon(x)=+1$ if $x>0$ and $-1$ if
$x<0$.  Commuting this exponential part of the $V_N$ vertex to the
right past all the other vertices gives the factor ${\rm exp}(-i\pi
k^2_N)$, where we have used momentum conservation.  Since this factor
is independent of the number of vertices $V_N$ was commuted through,
whereas the phase that $V_0(k_N,x)$ picks up will depend on how many
other $V_0$'s it commutes with, the only requirement consistent with
having non-zero scattering of arbitrary numbers of particles is that
$k^2_N\in 2\IZ$ and the $V_0(k_i,x)$ commute with each other.

Now, from the representation theory of the FSC algebra, only
world-sheet fields with $\IZ_3$ charge zero can be commuting
operators.  Combined with the condition that the $V_i$ vertices have
conformal dimension one, this implies tight restrictions on the
possible candidate states appearing in the scattering amplitudes.  In
particular, if the physical state we want to scatter is a D-module
highest-weight state $W^\pm_d$, the appropriate operators appearing in
(\ref{Aone}) would have to be the $q=0$ Virasoro primary descendents of
$W^\pm_d$ of conformal dimension $h_d+n+{2\over3}$.  The lowest level
such states are $V^{(\pm)}_d$, with dimension $h_d+{2\over3}$.  This
choice for the $V$ vertex in (\ref{Aprop}) implies the $L_0$ intercept
for D-module highest-weight states to be $h_d={1\over3}$ in order for
the total dimension of the vertex to be 1.  We will examine this
possibility below, and return to the case of S-module physical state
scattering later.

Consider scattering of D-module highest-weight states, where the
vertices in Eq.~(\ref{Aprop}) may correspond to the $V^{(\pm)}_d$
descendent states in a FSC D-module.  We can convert Eq.~(\ref{Aprop})
to a different ``picture'' involving the highest-weight states
$W^\pm_d$ using the general properties of D-modules.  Evaluate the
commutator
	\begin{equation}
	\label{gvcom} [G^\pm_r,V^{(+)}_d(w)]\equiv\oint_w {dz\over 2\pi i}
	z^{r+1/3} G^\pm(z) V^{(+)}_d(w)
	\end{equation}
 (where the integration
contour is around the point $w$) by inserting the $G^\pm(z)
V^{(+)}_d(w)$ OPE in (\ref{Dmod}) on the right-hand side, since it
involves only integer powers of $z-w$.  Setting $w=1$, one finds
	\begin{eqnarray}
   \label{gtcom} [G^\pm_r,V^{(+)}_d(1)] &=& \left( h_d + {c\over12} +
	\lambda^\pm\Lambda^\mp\right) \left\{ \left( r + {1\over3}
	\right) W^\pm_d(1) + {1\over 3h_d} \partial W^\pm_d(1) \right\}
	\nonumber\\ &&\ \ \mbox{} - \left(\Lambda^\pm -
	{1\over2}\lambda^\pm \right) {\widetilde W^\pm_d}(1).
	\end{eqnarray}
In
deriving this commutator, we have only used general properties of
D-modules.  For string scattering, though, we should set the dimension
of $W^\pm_d$ to $h_d={1\over3}$, since at this value of the intercept
$V^{(+)}_d$ has dimension one, as required by conformal invariance.
The commutator (\ref{gtcom}) dramatically simplifies at this value of
$h_d$:
	\begin{equation}
	\label{gfcom}
	[G^\pm_r,V^{(+)}_d(1)]=\left\{\left(r+{1\over3}\right)
	W^\alpha_d(1)+\partial W^\alpha_d(1)\right\}.
	\end{equation}
 Since
$W^\pm_d$ are Virsoro primaries of dimension $h_d$,
$[L_0,W^\pm_d(1)]=h_d W^\pm_d(1)+\partial W^\pm_d(1)$, and using this
in (\ref{gfcom}) with $h_d=1/3$ then gives
	\begin{equation}
	\label{gocom}
	[G^\pm_r,V^{(+)}_d(1)]=\left(L_0+r-{1\over3}\right)W^\pm_d(1)
	-W^\pm_d(1)\left(L_0-{1\over3}\right)
	\end{equation}
 for all $r\in
	\IZ/3$.

The crucial point for what follows is that at $h_d=1/3$ the
dimension-$(1\!+\!h_d)$ descendents $\widetilde W^\pm_d$ decoupled from
the commutator (\ref{gtcom}).  It is this unexpected decoupling at
precisely the physical value of the intercept which will allow us to
construct sensible tree scattering amplitudes from Eq.~(\ref{gocom}).
Note that the decoupling of $\widetilde W^\pm_d$ has no counterpart in
the analogous formulation of ordinary superstring scattering
amplitudes.  The occurence of this operator in the first place is due
to the non-linear structure of the FSC algebra, and its decoupling
appears as the result of an algebraic ``accident'' in this formulation
of fractional superstring scattering amplitudes.  Note also that the
decoupling does not occur for the $V^{(-)}_d$ descendent state.  As a
result, only $V^{(+)}_d$ will be a consistent choice for the vertices
appearing in the amplitude (\ref{Aprop}).

With this understanding, we interpret all the vertices in (\ref{Aprop})
as $V^{(+)}_d$ descendent states of D-module physical states $W^\pm_d$,
and we drop the $d$ subscript.  We can now replace the ``in'' state
$\ket{V^{(+)}_1}$ in (\ref{Aprop}) with a physical state using
$\ket{V^{(+)}}=G^-_{-2/3}\ket{W^+} + G^+_{-2/3}\ket{W^-}$ which follows
from Eq.~(\ref{Ddesc2}).  The $G^\pm_{-2/3}$ modes can be commuted to
the left using Eq.~(\ref{gocom}) as well as the relation
	\begin{equation}
	\label{propcom} G^\pm_r(L_0-a-r)^{-1}=(L_0-a)^{-1}G^\pm_r ,
	\end{equation}
following from Eq.~(\ref{vircom}).  Acting on the ``out'' state,
$\langle V^{(+)}|G^\pm_{-2/3}=\alpha\langle W^\pm|$, which is a
consequence of the third OPE in Eq.\ (\ref{Dmod}) with $h_d=1/3$.  The
extra insertions coming from the right-hand side of Eq.~(\ref{gocom})
vanish by a ``cancelled propagator'' argument, since setting $r=-2/3$
in Eq.~(\ref{gocom}) gives factors of $L_0-1$ and $L_0-1/3$ which
cancel the propagators to the left and right, respectively.  Tree
amplitudes with cancelled propagators are holomorphic in the Mandelstam
invariant of the cancelled propagator channel, and thus, by
analyticity, vanish if the amplitudes have Regge asymptotic behavior.
We will see in the next section that they do have this soft high energy
behavior.  The resulting form for the scattering amplitude is
	\begin{equation}
	\label{Apict2} {\cal A}_N= \langle W^+_N| V^{(+)}_{N-1}(1)
	\Delta\ldots\Delta V^{(+)}_2(1) \ket{W^-_1} + \langle W^-_N|
	V^{(+)}_{N-1}(1) \Delta\ldots\Delta V^{(+)}_2(1) \ket{W^+_1} ,
	\end{equation}
 where the propagator in this picture is
$\Delta=(L_0-{1\over3})^{-1}$.  The two terms appearing in
(\ref{Apict2}) are actually equal, as is easy to see using the
normalization of the $W^\pm_d$ states given in (\ref{Dint}).  In
particular, one can rewrite $\langle W^+_N|$ as
$(\Lambda^-)^{-1}\langle W^-_N|G^-_0$ and then commute the $G^-_0$ to
the right using (\ref{gocom}), (\ref{propcom}), and the cancelled
propagator argument until it acts on $\ket{W^-_1}$ to give
$\Lambda^-\ket{W^+_1}$.  Thus, the final form we find for the
scattering amplitude of D-module physical states is
	\begin{eqnarray}
   \label{Apict} {\cal A}_N &=& 2\langle W^+_N| V^{(+)}_{N-1}(1)
	\Delta\ldots\Delta V^{(+)}_2(1) \ket{W^-_1} \nonumber\\
	&=& 2\langle W^-_N| V^{(+)}_{N-1}(1)
        \Delta\ldots\Delta V^{(+)}_2(1) \ket{W^+_1} .
	\end{eqnarray}
These two forms for ${\cal A}_N$ along with the expression (\ref{Aprop})
in terms of $q=0$ vertices comprise three physically equivalent
``pictures'' for computing scattering amplitudes.  They are
clearly closely related to the $\IZ_3$ symmetry of the spin-4/3
FSC algebra.

Now we can investigate the crucial issue of spurious state decoupling
in our amplitudes.  If we start with physical states defined as
highest-weight vectors of FSC modules, will they scatter only to other
physical states?  For this to be true, only physical states must
contribute to residues of poles in amplitudes when an internal
propagator goes on-shell.  Suppose we fix the external momenta such
that some state $\ket{s}$ in the string Fock space at momentum
$\kappa=k_{M+1}+\cdots+k_N$ is on-shell:  $(L_0-1/3)\ket{s}=0$.  If we
factorize the amplitude in Eq.~(\ref{Apict}) by inserting a sum over a
complete set of states of momentum $\kappa$ at the propagator between
$V^{(+)}_{M+1}$ and $V^{(+)}_M$, then the $\ket{s}\langle s|$ term in
the sum will contribute a pole in momentum space.  The requirement of
spurious state decoupling is that if $\ket{s}$ is spurious, its
contribution to the residue of the pole should vanish:
	\begin{equation}
	\label{spstde} \langle s|V^{(+)}_M(1)\Delta\cdots\Delta
	V^{(+)}_2(1)\ket{W^-_1}=0 .
	\end{equation}
 To prove this, consider one
term, say $\langle\psi^-|G^-_r$ with $r>0$, in the presentation of
$\langle s|$ as a sum of descendent states, Eq.~(\ref{spur}), where
$\ket{\psi^-}$ must satisfy $(L_0+r-{1\over3})\ket{\psi^-} = 0$.
(The $G^+_r$ descendent pieces can be shown to decouple by the same
argument, and the $L_n$ pieces by a similar, simpler argument.)  The
$G^-_r$ mode can be commuted to the right in Eq.~(\ref{spstde}) using
Eqs.~(\ref{gocom}) and (\ref{propcom}).  The insertions coming from the
right-hand side of Eq.~(\ref{gocom}) again vanish by a cancelled
propagator argument.  Finally, the $G^-_r$ mode acting on the ``in''
state $\ket{W_1}$ vanishes by the physical state conditions
Eq.~(\ref{psc}), thus proving spurious state decoupling.

To examine whether similar considerations can give sensible scattering
amplitudes for S-module physical states, first of all note that the
intercept in this sector does not have to be the same as that of the
D-module physical states.  As discussed above, the conditions coming
from conformal invariance for an operator to represent a scattering
vertex in (\ref{Aprop}) are that it have $\IZ_3$ charge $q=0$ and
be Virasoro primary of conformal dimension 1.  In an S-module,
appropriate operators would have to be $W_s$ itself (of conformal
dimension $h_s$) or one of its Virasoro primary descendents of
dimension $h_s+n$ in order to satisfy the $q=0$ condition.  Choosing
the intercept $h_s=1$ implies that the $V_i$ in (\ref{Aprop}) be
identified with S-module highest-weight states $W_s$.  However, from
the S-module OPEs (\ref{Smod}) we can derive the relation analogous to
(\ref{gocom}):
	\begin{equation}
	\label{bgrcom} [G^\pm_r,W_s(1)]=V^\pm_s(1)\qquad\qquad{\rm
	for\ all}\ r\in\IZ/3.
	\end{equation}
 Because this commutator does not
have the factors of $L_0$ on the right-hand side similar to those that
appeared in (\ref{gocom}), there can be no cancelled propagator
argument to remove the right-hand side of (\ref{bgrcom}) when used in
evaluating the residue of a spurious state pole, and thus the spurious
state decoupling proof fails.  One might wish to consider instead the
S-module descendent $\widetilde W_s$ with dimension $h_s+1$ by choosing
the S-sector intercept $h_s=0$.  {}From the last OPE in
Eq.\ (\ref{Smod}) follows a commutator similar to that of
(\ref{gtcom}).  For the picture-changing and spurious state decoupling
arguments to go through, though, the contibutions of the dimension
$h_s+{4\over3}$ descendents ${\widetilde V}^\pm_s$ and
${\widetilde{\widetilde V}}^\pm_s$ must decouple from the commutator
when $h_s=0$.  One finds that ${\widetilde V}^\pm_s$ does not
decouple.  Bosonic sector descendents at higher levels also do not seem
to work since they also are created from the $W_s$ primary by quadratic
or higher combinations of current modes.  This makes a picture-changing
argument of the type used to relate (\ref{Aprop}) to (\ref{Apict})
problematical.

To summarize, our prescription for dual N-point tree amplitudes
satisfying spurious state decoupling is
	\begin{equation}
	\label{Asum} {\cal A}_N=\langle W^+_d|V^{(+)}_d(1) {1\over
	L_0-{1\over3}} V^{(+)}_d(1)\ldots{1\over L_0-{1\over3}}
	V^{(+)}_d\ket{W^-_d}
	\end{equation}
 where $W^\pm_d$ are S-module physical
states which are required to have  $L_0$ intercept
	\begin{equation}
	\label{gopsc} L_0\ket{W^\pm_d} = {1\over3}\ket{W^\pm_d},
	\end{equation}
 and
their relative normalizations are fixed by
	\begin{equation}
	\label{normal} G^\pm_0\ket{W^\pm_d} = \Lambda^\pm\ket{W^\mp_d},
	\end{equation}
 where $\Lambda^+\Lambda^-= {1\over3} -{c\over24}$ is fixed by
associativity of the FSC algebra, and we choose $\Lambda^+=\sqrt{\vert
{1\over3} - {c\over24}\vert}$.  With this convention the $V^{(+)}_d(x)$
fields are the $q=0$ descendent states
	\begin{equation}
	 \ket{V^{(+)}_d} = G^+_{-2/3} \ket{W^-_d} + G^-_{-2/3}
	\ket{W^+_d}.
	\end{equation}
 (Each state or vertex in the amplitude can
correspond to a different physical state, of course.)

This prescription can be extended to include two S-module physical
states by simply replacing the ``in'' and ``out'' $W^\pm_d$ states in
Eq.~(\ref{Apict}) with S-module states $W_s$:
	\begin{equation}
	\label{Absum} {\cal A}_N=\langle W_s|V^{(+)}_d(1) {1\over L_0-h_s}
	\ldots{1\over L_0-h_s} V^{(+)}_d(1)\ket{W_s}.
	\end{equation}
 The argument
for spurious state decoupling then goes through unchanged.  However,
since there is no appropriate dimension-1 commuting vertex in the
S-module to play the role of the $V^{(+)}_d$ vertices, we cannot prove
cyclic symmetry of the amplitudes with two S-module states, nor can we
extend the prescription to include scattering of three or more S-module
states.  This situation is closely analogous to what happens in the old
covariant formalism in the ordinary superstring.  There, dual
amplitudes with spurious state decoupling can be formulated for
scattering of Neveu-Schwarz sector states, and can only be extended to
include two Ramond-sector states as the ``in'' and ``out'' states in
the correlator, thus losing manifest cyclic symmetry.  So presumably,
just as in the Ramond sector of the superstring, our inability to
incorporate more than two S-module physical states in our scattering
prescription means that there is a nontrivial contribution to S-module
scattering amplitudes coming from the ``fractional ghost'' fields on
the world-sheet.

Note, however, that upon factorizing the D-module scattering amplitude
in Eq.~(\ref{Asum}) on any propagator, we can never obtain an S-module
intermediate state.  The reason is simply that the D-module $W^-_d$
``in'' state has $\IZ_3$ charge $q=-1$ and their descendent
$V^{(+)}_d$ vertices have charge $q=0$.  By conservation of $\IZ_3$
charge, only $q=-1$ intermediate states can contribute, whereas the
S-module physical states $W_s$ have $q=0$.  This means, for example,
that
	\begin{equation}
	\label{Aselect} \langle W_s|V^{(+)}_d(1) \ket{W^-_d}=0.
	\end{equation}
 This
selection rule makes it consistent at tree level to drop the S-module
physical states altogether, a desirable feature if it turns out that
the S-module physical states include tachyons.
This projection is closely analogous to the GSO projection in
the Neveu-Schwarz sector of the ordinary superstring \cite{GSO}.

\section{\label{sIV}A three-dimensional fractional superstring}

Any string propagating in $D$ flat space-time dimensions will be
described by a world-sheet CFT which includes $D$ massless scalar
fields $X^\mu(\sigma,\tau)$.  These fields are interpreted as giving
the space-time coordinates $X^\mu$ of the point $(\sigma,\tau)$ on the
string world-sheet.  The idea behind the construction of the spin-4/3
fractional superstring is to require that the world-sheet CFT also
include a set of fields $\epsilon^\pm_\mu(\sigma,\tau)$ of right-moving
(holomorphic) conformal dimension 1/3, transforming as vectors under
space-time Lorentz transformations.  In addition, we demand that there
exist currents $G^\pm(\sigma,\tau)$ of conformal dimension 4/3 on the
world-sheet of the form $G^\pm(\sigma,\tau)=\epsilon^\pm_\mu \partial
X^\mu +\ldots$, obeying the FSC algebra (\ref{fsca}).  As described in
the last section, this algebra organizes the world-sheet CFT, allowing
us to identify vertex operators corresponding to physical states.

In this section we construct an example of a CFT satisfying these
requirements, and having a three-dimensional space-time
interpretation.  We compute a few low-lying states in the spectrum of
this fractional superstring and calculate some of their scattering
amplitudes.  The CFT in question is particularly simple, being
constructed from five free massless scalar fields on the world-sheet.
In describing this theory, we only write the right-moving parts of the
world-sheet fields, {\it i.e.}\ those holomorphic in $z=\tau+i\sigma$.
By itself this is suitable for describing an open string; if matched
with an appropriate left-moving theory it will describe a closed or
heterotic string.

Since we construct this CFT from five free massless scalars, it will
have central charge $c=5$.  Three of the scalars are just coordinate
boson fields $X^\mu(z)$, $\mu=0,1,2$, with the standard operator
products
	\begin{equation}
	\label{Xope} X^\mu(z) X^\nu(w)=-g^{\mu\nu}\ln(z-w)\ ,
	\end{equation}
 where
$g^{\mu\nu}$ is the three-dimensional Minkowski metric with signature
$(-++)$.  The stress-energy tensor for these fields is given by
	\begin{equation}
	\label{Xstress} T_X = -{1\over2} g_{\mu\nu} \partial X^\mu
	\partial X^\nu .
	\end{equation}
 This $X^\mu$ CFT has a global $so(2,1)$
Lorentz symmetry, generated by the charges
	\begin{equation}
	\label{2.3} M_X^{\mu\nu}=\oint{dz\over 2\pi i}\left(X^\mu\partial
	X^\nu - X^\nu\partial X^\mu\right)\ .
	\end{equation}
 We assign
hermiticities and choose conventions so that $(X^\mu)^\dagger =X_\mu$
and $(\partial X^\mu)^\dagger = -\partial X_\mu$.

The remaining two fields, $\varphi^i(z)$, $i=1,2$, describe a map from
the string world-sheet to a torus.  In a basis in which the target
space $\varphi^i$ boundary conditions are diagonalized,
	\begin{equation}
	\label{2.4} \varphi^i(z)=\varphi^i(z)+2\pi\ ,
	\end{equation}
 their operator
product expansion (OPE) is
	\begin{equation}
	\label{phiope} \varphi^i(z)\varphi^j(w)=-g^{ij}\ln(z-w)\ ,
	\end{equation}
where
	\begin{equation}
	\label{2.6} g^{ij}={1\over3}\pmatrix{\hfill 2&-1\cr -1&\hfill
	2\cr}\ .
	\end{equation}
 Alternatively, we could have chosen different
linear combinations of the $\varphi^i$, say $\tilde\varphi^i$, which
have the standard OPEs $\tilde\varphi^i(z)\tilde\varphi^j(w) =
-\delta^{ij}\ln(z-w)$, but which would then have boundary conditions
more complicated than those in (\ref{2.4}).  Introduce two pairs of
vectors ${\bf e}^i$ and ${\bf e}_i$ satisfying ${\bf e}^i\cdot{\bf e}^j
= g^{ij}$, ${\bf e}_i\cdot{\bf e}_j = g_{ij}$ and ${\bf e}^i\cdot{\bf
e}_j = \delta^i_j$, where $g_{ij}$ is the matrix inverse of $g^{ij}$,
and define the vector-valued field $\mbox{\boldmath$\varphi$} =
\varphi^i{\bf e}_i$.  The stress-energy tensor for scalars obeying the
OPEs (\ref{phiope}) is
	\begin{equation}
	\label{2.8} T_\varphi = -{1\over2}\partial
	\mbox{\boldmath$\varphi$} \cdot \partial
	\mbox{\boldmath$\varphi$} =
	-(\partial\varphi^1)^2-(\partial\varphi^2)^2
	-(\partial\varphi^1)(\partial\varphi^2) .
	\end{equation}

The {\boldmath$\varphi$}\ CFT also has a global $so(2,1)$ Lorentz
symmetry, and there are two dimension-1/3 fields, $\epsilon^+_\mu$ and
$\epsilon_\mu^-$, which transform as vectors under this symmetry. To
see this, consider the vertex operators
	\begin{equation}
	\label{2.7} V_{\bf m}=c({\bf m}){\rm exp}\{i{\bf m}\cdot
	\mbox{\boldmath$\varphi$}\}
	\end{equation}
 where ${\bf m}=m_i{\bf e}^i$
and $c({\bf m})$ is an appropriate cocylce operator, described in more
detail in \ref{appA}.  Because of the boundary conditions (\ref{2.4}),
these are well-defined fields for integer $m_i$.  The $V_{\bf m}$
vertex operators are Virasoro primary fields of conformal dimension
	\begin{equation}
	\label{2.9} h(V_{\bf m})={1\over2}{\bf m}\cdot{\bf m}=
	{1\over3}\left(m_1^2+m_2^2-m_1m_2\right) .
	\end{equation}
 It follows that
all vertex operators have dimensions either an integer or an integer
plus 1/3.  Note also that $m_1+m_2$ is (mod 3) just the $\IZ_3$
charge of the $V_{\bf m}$ operator.  The ``momenta'' ${\bf m}$ of the
vertex operators $V_{\bf m}$ take values in a triangular lattice, as
shown in Fig.~4.  This lattice is actually the $su(3)$ weight lattice
($g_{ij}$ is equivalent to the $su(3)$ Cartan matrix), and thus the
{\boldmath$\varphi$}\ CFT is the standard free boson realization of the
$su(3)_1$ Wess-Zumino-Witten model \cite{FK}.  An $so(3)$ current
algebra arises because $so(3)_2$ is conformally embedded in $su(3)_1$,
corresponding to the embedding of $so(3)$ as a non-regular subalgebra
of $su(3)$.  The difference between $so(3)$ and $so(2,1)$ is just a
matter of the appropriate choice of cocycles.  With these cocycles, the
vertex operators satisfy the basic operator product expansion
	\begin{equation}
	\label{2.9.5} V_{\bf m}(z) V_{\bf n}(w) = (-1)^{m_2n_1}(z-w)^{\bf
	m\cdot n} V_{\bf m+n}(w) + \ldots .
	\end{equation}

More concretely, consider the triplet of dimension-one fields, $U_\mu$,
	\begin{eqnarray}
   \label{2.10} U_0 &=& V_{(1,-1)} + V_{(-1,1)}  \nonumber\\ U_1 &=&
	V_{(1,2)}  + V_{(-1,-2)} \nonumber\\ U_2 &=& V_{(-2,-1)}  +
	V_{(2,1)}
	\end{eqnarray}
where we denote the vertex operator $V_{\bf m}$
by the components of ${\bf m}=m_i{\bf e}^i$ as $V_{(m_1,m_2)}$.  The
$U_\mu$ generate the $so(2,1)_2$ Kac-Moody current algebra
	\begin{equation}
	\label{2.11} U^\mu(z)U^\nu(w)={2g^{\mu\nu}\over(z-w)^2}
	+{\varepsilon^{\mu\nu\rho} U_\rho(w)\over(z-w)}+\ldots\ ,
	\end{equation}
where $\varepsilon^{\mu\nu\rho}$ is the completely antisymmetric tensor
density in three dimensions normalized by $\varepsilon^{012}=1$.  The
zero-modes of these currents,
	\begin{equation}
	\label{2.12} M_\varphi^{\mu\nu}=\oint{dz\over 2\pi
	i}\varepsilon^{\mu\nu\rho} U_{\rho}(z) ,
	\end{equation}
 generate the
global $so(2,1)$ Lorentz rotations.

All the fields in the {\boldmath$\varphi$}\ CFT can be organized in
$so(2,1)$ representations.  For example, some of the simplest Virasoro
primary fields in the {\boldmath$\varphi$}\ CFT are the $so(2,1)$
vector fields $\epsilon^+_\mu$ and $\epsilon^-_\mu$ of conformal
dimension 1/3 and the $so(2,1)$ scalars $s^+$ and $s^-$ of dimension
4/3, given by
	\begin{eqnarray}
   \label{covflds} \epsilon^+_\mu &=& \left(V_{(-1,-1)}\,,\,
	V_{(1,0)}\,,\, V_{(0,1)}  \right) \nonumber\\ \epsilon^-_\mu
	&=& \left( V_{(1,1)}\,,\, V_{(-1,0)}\,,\, V_{(0,-1)} \right)
	\nonumber\\ s^+ &=& {1\over3} \left[
	V_{(2,2)}+V_{(-2,0)}+V_{(0,-2)} \right] \nonumber\\ s^- &=&
	{1\over3} \left[ V_{(-2,-2)}+V_{(2,0)}+V_{(0,2)} \right]
	\end{eqnarray}
The other Virasoro primary fields up to dimension 4/3 are the $U_\mu$
and a symmetric-traceless $W_{\mu\nu}$ both of dimension 1, and a pair
of symmetic-traceless dimension 4/3 fields $t^+_{\mu\nu}$ and
$t^-_{\mu\nu}$. (The precise vertex operator definitions of these
fields are given in \ref{appA}.)  There are, of course, an infinite
tower of Virasoro primary fields of higher dimension on the
world-sheet.  The dimension-1/3 vector fields, $\epsilon^+_\mu$ and
$\epsilon^-_\mu$, are the analogs of the dimension-1/2 fermion fields
$\psi^\mu$ of the ordinary ten-dimensional superstring. Likewise, the
dimension-1 adjoint field $\varepsilon^{\mu\nu\rho} U_\rho$ is the
analog of the dimension-1 $\psi^\mu\psi^\nu$ superstring field.  On the
other hand, the dimension-4/3 scalars, $s^\pm$, and the dimension-1 and
-4/3 spin-2 fields, $W_{\mu\nu}$ and $t^\pm_{\mu\nu}$, have no
superstring analogs.  It is a straight-forward exercise to work out the
OPEs satisfied by the above fields using the free boson operator
product (\ref{phiope}).  The results are collected in Appendix A, where
attention has also been paid to the cocycle algebra, needed to get the
signs right.

We now construct the fractional supercurrents $G^\pm$.  This current
must be a dimension-4/3 Virasoro primary field, and a scalar with
respect to the global $so(2,1)$ Lorentz symmetry generated by
$M^{\mu\nu} = M_X^{\mu\nu}+M_\varphi^{\mu\nu}$.  In addition, it should
be invariant under translations along the $X^\mu$ directions, generated
by the momentum $P^\mu=i\oint{dz\over2\pi i}\partial X^\mu$, which
together with $M^{\mu\nu}$ generates the full three-dimensional
Poincar\'{e} group. This implies that $G^\pm$ can only depend on
derivatives of $X^\mu$ and not on the $e^{ik\cdot X}$ vertex
operators.  There are clearly only four fields which obey these
requirements:  $\epsilon^\pm_\mu\partial X^\mu$ and $s^\pm$.  The
coefficients with which these four fields contribute to $G^\pm$ are
fixed by requiring that the OPEs of $G^\pm$ with themselves close only
on the stress-energy tensor $T=T_X+T_\varphi$ and $G^\pm$ among its
singular terms.  Using the OPEs tabulated in Appendix A, the fractional
supercurrent is found to be
	\begin{eqnarray}
   \label{Grepd} G^+&=&{1\over\sqrt2}\left(\epsilon^+\cdot\partial X
	-{3\over2}s^+\right) ,\nonumber\\
	G^-&=&{1\over\sqrt2}\left(-\epsilon^-\cdot\partial X
	-{3\over2}s^-\right) .
	\end{eqnarray}
In fact, the coefficents of the
terms in $G^\pm$ are overconstrained by the condition that a fractional
superconformal algebra closing only on $G^\pm$ and $T$ should exist.
The existence of the solution (\ref{Grepd}) is an indication that we
have in fact chosen a special world-sheet CFT: the generic CFT with a
global Lorentz symmetry and dimension-1/3 vector fields would not have
a fractional superconformal symmetry.  $G^\pm$ and $T$ satisfy the FSC
operator product algebra (\ref{fsca},\ref{lamb}) with $c=5$.

We should comment on the uniqueness of the expression (\ref{Grepd}) for
the fractional supercurrent.  First, note that the replacement $V_{\bf
m} \rightarrow \widetilde V_{\bf m} = \beta^{m_1}\gamma^{m_2} V_{\bf
m}$ is a symmetry of the basic operator product (\ref{2.9.5}), where
$\beta$ and $\gamma$ can be arbitrary complex numbers.  (This can be
thought of as the result of a complex shift in the origin of the
$\varphi^j$ boson fields.)  Thus, making this replacement in the
expressions (\ref{Grepd}) for $G^\pm$ and then re-expressing the
$\widetilde V_{\bf m}$'s in terms of $\beta$, $\gamma$ and the old
$V_{\bf m}$'s will give new expressions for $G^\pm$ which will
automatically obey the same operator product algebra.  Similarly, the
$X^\mu$ OPEs are preserved under the replacement
$X^\mu\rightarrow\Lambda^\mu_\nu X^\nu$ where $\Lambda^\mu_\nu$ is any
$so(2,1)$ rotation, and so any such repalcement will preserve the FSC
algebra OPEs.  However, such transformations will not, in general,
preserve $so(2,1)$ invariance.  Indeed, it is easy to check that there
are only six such transformations that {\it do} preserve the space-time
Lorentz invariance.  The resulting six solutions for $G^\pm$ are
	\begin{eqnarray}
	 G^\pm &\rightarrow & \omega^{\pm q} G^\pm ,\nonumber\\ {\rm
	or}\quad G^\pm &\rightarrow & \omega^{\pm q} \widetilde G^\pm ,
	\end{eqnarray}
where $\omega=e^{2\pi i/3}$, $q\in\IZ_3$ and
$\widetilde G^\pm$ are given by
	\begin{eqnarray}
   \label{Grepspltil} \widetilde
	G^+&=&{1\over\sqrt2}\left(-\epsilon^+\cdot\partial X
	-{3\over2}s^+\right) ,\nonumber\\ \widetilde
	G^-&=&{1\over\sqrt2}\left(\epsilon^-\cdot\partial X
	-{3\over2}s^-\right) ,
	\end{eqnarray}
which differs from the solution in
(\ref{Grepd}) by a sign change in the $\epsilon^\pm\cdot\partial X$
terms.  The existence of these six solutions is a consequence of the
$\IZ_2\times S_3$ automorphism group of the $c=5$ CFT generated by
$X^\mu \rightarrow -X^\mu$, $V_{\bf m} \rightarrow {\rm exp}\{2\pi
i(m_1+m_2)/3\}V_{\bf m}$, and $V_{\bf m} \rightarrow V_{-{\bf m}}$,
which leaves the $so(2,1)$ generators invariant.
These six solutions for $G^\pm$
give rise to equivalent representation
theories and thus it is immaterial which of them
we choose to be the generators of our physical state
conditions.

\subsection{Physical state conditions for general vertex operators}

In this subsection we will set up an efficient formalism for computing
the action of the modes of the currents $G^\pm$ and $T$ on a general
state.  This enables one to determine, in principle, the physical
states at arbitrarily high levels.  The method we use involves
generalized commutation relations similar to those derived in Section
\ref{sII}.  The reader unfamilar with generalized commutators may
safely skip to the next subsection where the lowest-lying physical
states and their scattering amplitudes are computed using only the
(free boson) OPEs collected in \ref{appA}.

It is a complicated problem to identify the Virasoro primary $so(2,1)$
covariant combinations of fields of high dimension in the
{\boldmath$\varphi$} CFT.  In addition, to compute the action of the
physical state conditions on these fields, one must calculate their
OPEs with the $G^\pm$ currents, which can be a lengthy procedure.   A
way around this is to express all the states in the
{\boldmath$\varphi$} CFT in terms of the modes of the
$\epsilon^\pm_\mu$ fields.  In this basis $so(2,1)$ covariance is
manifest.  Also, since the modes of the currents $G^\pm$ and $T$ can be
written in terms of $\epsilon^\pm_\mu$ modes, all that is needed to
compute the physical state conditions on a given state are the
generalized commutation relations of the $\epsilon^\pm_\mu$ modes.

These generalized commutation relations (GCRs) can be derived from the
$\epsilon^\pm_\mu$ OPEs given in \ref{appA} in the same way that the
$G^\pm$ GCRs were derived in Section \ref{sII}.  In particular, picking
up just the first term of the $\epsilon^\pm_\mu\epsilon^\pm_\nu$ OPE
gives
	\begin{equation}
	\label{epgcrpp}
	\sum_{\ell=0}^\infty C^{(-2/3)}_\ell \left\{
	\epsilon^{\mu\pm}_{m - \ell \pm {q\over3}}
	\epsilon^{\nu\pm}_{n + \ell + {2 \pm q\over3}} -
	\epsilon^{\nu\pm}_{n - \ell \pm {q\over3}}
	\epsilon^{\mu\pm}_{m + \ell + {2 \pm q\over3}}
	\right\} = {\varepsilon^{\mu\nu}}_\rho
	\epsilon^{\rho\mp}_{m + n + {2 \pm 2q \over 3}}
	\end{equation}
when acting on any state with $\IZ_3$ charge $q$.  (The rules for the
allowed modings of the $\epsilon^\pm$ fields are the same as that for
the $G^\pm$ currents summarized in Fig.\ 2.)  The binomial coefficient
$C^{(\alpha)}_\ell$ is given in Eq.\ (\ref{cexp}).  Picking up only the
leading term of the $\epsilon^\pm_\mu \epsilon^\mp_\nu$ OPE gives
	\begin{equation}
	\label{epgcrpm}
	\sum_{\ell=0}^\infty C^{(-1/3)}_\ell \left\{
	\epsilon^{\mu\pm}_{m - \ell + {1 \pm q\over3}}
	\epsilon^{\nu\mp}_{n + \ell + {2 \mp q\over3}} +
	\epsilon^{\nu\mp}_{n - \ell + {1 \mp q\over3}}
	\epsilon^{\mu\pm}_{m + \ell + {2 \pm q\over3}}
	\right\} = g^{\mu\nu} \delta_{m + n + 1} .
	\end{equation}
Any state in the {\boldmath$\varphi$} CFT can be written as a
polynomial in the $\epsilon^\pm$ creation modes acting on the vacuum.
The GCRs (\ref{epgcrpp}) and (\ref{epgcrpm}) are sufficient to reduce
any set of such states to a linearly independent basis.

The current modes can be expressed in terms of $\epsilon^\pm$ modes as
follows.  Since $\epsilon^\pm \cdot \epsilon^\pm = 3 z^{2/3} s^\mp +
\ldots$, one derives
	\begin{eqnarray}
   \label{stoep}
	s^\pm_{2m - {1 \mp q\over3}} &=&
	{1\over3} \sum_{\ell=0}^\infty C^{(-5/3)}_\ell \left\{
	\epsilon^\mp_{m - \ell -1 \mp {q\over3}} \cdot
	\epsilon^\mp_{m + \ell + {2 \pm 2q\over3}} +
	\epsilon^\mp_{m - \ell -1 \pm {2q\over3}} \cdot
	\epsilon^\mp_{m + \ell + {2 \mp q\over3}}
	\right\} , \nonumber\\
	s^\pm_{2m + 1 - {1 \mp q\over3}} &=&
	{1\over3} \sum_{\ell=0}^\infty C^{(-5/3)}_\ell \left\{
	\epsilon^\mp_{m - \ell  \mp {q\over3}} \cdot
	\epsilon^\mp_{m + \ell  + {2 \pm 2q\over3}} +
	\epsilon^\mp_{m - \ell - 1 \pm {2q\over3}} \cdot
	\epsilon^\mp_{m + \ell + 1 + {2 \mp q\over3}}
	\right\} ,
	\end{eqnarray}
and since $\epsilon^\pm \cdot \epsilon^\mp = 3 z^{-2/3} + z^{4/3}
T_\varphi + \ldots$,
	\begin{eqnarray}
   \label{ttoep}
	L^\varphi_{2m} &=&
	\sum_{\ell=0}^\infty C^{(-7/3)}_\ell \left\{
	\epsilon^+_{m - \ell - {2 - q\over3}} \cdot
	\epsilon^-_{m + \ell + {2 - q\over3}} +
	\epsilon^-_{m - \ell - {5 + q\over3}} \cdot
	\epsilon^+_{m + \ell + {5 + q\over3}}
	\right\} - {q(q+3)\over6} \delta_m , \nonumber\\
	L^\varphi_{2m + 1} &=&
	\sum_{\ell=0}^\infty C^{(-7/3)}_\ell \left\{
	\epsilon^+_{m - \ell - {2 - q\over3}} \cdot
	\epsilon^-_{m + \ell + {5 - q\over3}} +
	\epsilon^-_{m - \ell - {2 + q\over3}} \cdot
	\epsilon^+_{m + \ell + {5 + q\over3}}
	\right\} ,
	\end{eqnarray}
where $L^\varphi_n$ are the modes of $T_\varphi$.  Introduce also the
usual mode expansion for the $X^\mu$ fields:
	\begin{equation}
	 X^\mu(z) = x^\mu - i\alpha_0^\mu {\rm ln}(z) +
	i\sum_{n\neq0} {1\over n} \alpha^\mu_n z^{-n}  ,
	\end{equation}
 satisfying
the standard commutation relations $[x^\mu, \alpha_0^\nu] =
ig^{\mu\nu}$ and $[\alpha^\mu_m, \alpha^\nu_n] = m\delta_{m+n}
g^{\mu\nu}$.   Then, from the expressions for the currents $G^\pm$ and
$T$ worked out earlier in this section, one finds
	\begin{eqnarray}
   \label{currmodestoep} L_n &=& {1\over 2}
	\sum_{\ell=-\infty}^\infty \alpha_{m-\ell} \cdot \alpha_\ell +
	L^\varphi_m , \nonumber \\ G^\pm_r &=& \mp {i\over\sqrt2}
	\sum_{\ell=-\infty}^\infty \epsilon^\pm_{r-\ell} \cdot
	\alpha_\ell - {3\over 2\sqrt2} s^\pm_r ,
	\end{eqnarray}
which, using
(\ref{stoep}) and (\ref{ttoep}) gives the current modes solely in terms
of $\epsilon^\pm$ and $\alpha$ modes.

\subsection{Simple vertex operators}

The fields of our $c=5$ CFT are organized into highest-weight modules
of the fractional superconformal algebra.  We refer to the two fields
of lowest conformal dimension in a module as a ``fractional
superconformal multiplet''.  The modules are characterized by the
dimensions and $\IZ_3$ charges of the multiplet fields.  These and
other properties of the fractional superconformal modules were derived
in a general way, independent of any particular CFT representation of
the fractional superconformal algebra, in Section~\ref{sII}.  For the
purposes of this section, we just illustrate these facts by considering
two of the simplest vertex operators in the $c=5$ theory,
	\begin{eqnarray}
   \label{2.17} W_s &=& e^{ik\cdot X} ,\nonumber\\ W^\pm_d &=&
	\zeta^{\pm\mu}\epsilon^\pm_\mu e^{ik\cdot X} ,
	\end{eqnarray}
which
describe, respectively, scalar and vector particles in space-time as
shown by their $so(2,1)$ transformation properties.  The
$\zeta^\pm_\mu$ coefficients in $W_d$ are polarization vectors, and the
$k^\mu$ are interpreted as space-time momenta.  Both vertices are
Virasoro primary fields of conformal dimensions $h(W_s) = {1\over2}k^2$
and $h(W^\pm_d) = {1\over2}k^2 + {1\over3}$.

Before deriving the properties of these vertex operators in detail, let
us summarize the results relevant for the computation of scattering
amplitudes as discussed in Section \ref{sIII}.  The properly normalized
$W^\pm_d$ vertices satisfying the physical state conditions are
	\begin{equation}
	 W^\pm_d=\left( \pm\xi\cdot\epsilon^\pm
	-ik\wedge\xi\cdot\epsilon^\pm \right) e^{ik\cdot X} ,
	\end{equation}
where $(A\wedge B)^\mu=\varepsilon^{\mu\nu\rho}A_\nu B_\rho$, the
polarization $\xi^\mu$ is transverse $\xi\cdot k=0$, and the state is
massless $k^2=0$.  The first FSC algebra descendent of this state is
	\begin{equation}
	 V^{(+)}_d = -\sqrt{2}\Bigl[\xi\cdot\partial X -
	ik\wedge\xi\cdot U + k^\mu(k\wedge\xi)^\nu W_{\mu\nu} \Bigr]
	{\rm e}^{ik\cdot X} .
	\end{equation}
 $W^\pm_d$ and $V^{(+)}_d$ are the
vertices appropriate for computing scattering amplitudes using the
prescription (\ref{Apict}) derived in Section \ref{sIII}.

The operator product algebra of the fractional superconformal currents
$G^\pm$ with $W_s$ is easily worked out:
	\begin{equation}
	\label{2.18}
	G^\pm(z)W_s(w)={V^\pm_s(w)\over(z-w)}+regular\ terms\ ,
	\end{equation}
where
	\begin{equation}
	\label{2.19} V^\pm_s = \mp{i\over\sqrt2} k\cdot\epsilon^\pm
	e^{ik\cdot X}
	\end{equation}
 are Virasoro primary fields of dimension
${1\over2}k^2+{1\over3}$.  We should think of $V^\pm_s$ as forming a
``fractional supermultiplet'' with $W_s$.  Note that no cuts occur in
the OPE (\ref{2.18}) reflecting the fact that $W_s$ is single-valued
with respect to the currents $T$ and $G^\pm$.  We describe this
situation by saying that $W_s$ is a ``bosonic'' field on the
world-sheet.  Computing, say,  the $G^\pm V^\mp_s$ OPE,
	\begin{equation}
	\label{2.20} G^\pm(z) V^\mp_s(w) = {{1\over2}k^2 W_s(w) \over
	(z-w)^{5/3}} + {{1\over2}\partial W_s(w) \pm \widetilde W_s(w)
	\over (z-w)^{2/3}} + \ldots ,
	\end{equation}
 we see that it closes back on
$W_s$, along with the higher-dimension Virasoro primary operator
$\widetilde W_s$ whose form will not be important to us.  The
fractional powers of $(z-w)$ appearing in (\ref{2.20}) reflect the
``fractional statistics'' of $V^\pm_s$ on the world-sheet.
Summarizing, we have found that the pair of fields $(W_s,V^\pm_s)$
belong to a fractional superconformal multiplet with conformal
dimensions $({1\over2}k^2,{1\over2}k^2+{1\over3})$ and with world-sheet
statistics {\it (bosonic, fractional)}.

We can perform a similar analysis for the $W^\pm_d$ field in
(\ref{2.18}).  The first few terms of the $G^\pm W^\pm_d$ OPEs are
	\begin{eqnarray}
   \label{2.21} G^\mp(z)W^\pm_d(w) &=& {\pm i k\cdot\zeta^\pm \over
	\sqrt2 (z-w)^{5/3}} e^{ik\cdot X}(w) + \ldots\nonumber\\
	G^\pm(z)W^\pm_d(w) &=& { (-\zeta^\pm \mp 2ik\wedge\zeta^\pm)
	\cdot \epsilon^\mp \over 2\sqrt2 (z-w)^{4/3}} e^{ik\cdot X}(w)
	+ \ldots
	\end{eqnarray}
Since the operator $e^{ik\cdot X}$ in the first
term of the first OPE has lower conformal dimension than $W^\pm_d$, it
follows that $W^\pm_d$ is, in general, not a primary fractional
superconformal field.  In fact, if the polarization vectors take the
form $\zeta^\pm_\mu \sim k_\mu$, then we recognize $W^\pm_d$ as the
$V^\pm_s$ member of the $W_s$ fractional supermultiplet.  For $W^\pm_d$
to be the highest member of its own fractional supermultiplet, we must
require the coefficient of the $(z-w)^{-5/3}$ term in (\ref{2.21}) to
vanish:
	\begin{equation}
	\label{2.22} k\cdot\zeta^\pm = 0 .
	\end{equation}
 In addition, to normalize
the $W^\pm_d$ vertices according to the prescription (\ref{normal})
used to define scattering amplitudes in Section \ref{sIII}, we demand
that the coefficients of the $(z-w)^{-4/3}$ terms satisfy
	\begin{equation}
	\label{2.23} -\zeta^\pm_\mu \mp 2i(k\wedge\zeta^\pm)_\mu =
	2\sqrt{2}\Lambda \zeta^\mp_\mu ,
	\end{equation}
 where $2\sqrt{2} \Lambda =
\sqrt{4k^2+1}$.  (Actually, $\Lambda$ is fixed to this value by
consistency of these equations, and does not, therefore, represent an
independent requirement on $\zeta^\pm_\mu$.) The solution to
(\ref{2.22}) and (\ref{2.23}) can be expressed in terms of a single
transverse polarization vector $\xi^\mu$: $\xi\cdot k = 0$ and
	\begin{equation}
	\label{2.24} \zeta^\pm_\mu = \pm\xi_\mu - { 2i(k\wedge\xi)_\mu
	\over 1+\sqrt{4k^2+1} } .
	\end{equation}

The $G^\pm W^\pm_d$ OPEs can then be computed:
	\begin{eqnarray}
   \label{2.26} G^\pm(z) W^\pm_d(w) &=& \left( \sqrt{4k^2+1} \over
	2\sqrt2 \right) {W^\pm_d(w) \over (z-w)^{4/3}} + \ldots
	\nonumber\\ G^\mp(z) W^\pm_d(w) &=& {1\over 2} {\left\{
	V^{(+)}_d(w) \mp V^{(-)}_d(w) \right\} \over (z-w)^{2/3}} +
	\ldots \nonumber\\ G^\pm(z) V^{(+)}_d(w) &=& \left({1\over2}k^2
	+ {3\over4} + {1\over4}\sqrt{4k^2+1}\right) {1\over(z-w)^2}
	\Biggl\{W^\pm_d(w) \\ &&\quad\mbox{} +
	{2(z-w)\over3k^2+2} \partial W^\pm_d(w)\Biggr\} -
	\left({\sqrt{4k^2+1} - 1 \over 2\sqrt2}\right) {\widetilde
	W^\pm_d(w)\over(z-w)} + \ldots \nonumber\\ G^\pm(z)
	V^{(-)}_d(w) &=& \mp \left({1\over2}k^2 + {3\over4} -
	{1\over4}\sqrt{4k^2+1}\right) {1\over(z-w)^2}
	\Biggl\{W^\pm_d(w) \nonumber\\ &&\quad\mbox{} +
	{2(z-w)\over3k^2+2} \partial W^\pm_d(w)\Biggr\} \pm
	\left({\sqrt{4k^2+1} + 1 \over 2\sqrt2}\right) {\widetilde
	W^\pm_d(w)\over(z-w)} + \ldots \nonumber
	\end{eqnarray}
where we have also written
the OPEs of $G^\pm$ with the $V^{(\pm)}_d$ descendents of $W^\pm_d$.
These OPEs are, of course, special cases of the D-module result
(\ref{Dmod}) derived in a more general context in Section \ref{sII},
and shown to be crucial for spurious state decoupling in tree-level
scattering amplitudes in Section \ref{sIII}.  The form of the
$V^{(\pm)}_d$ and $\widetilde W^\pm_d$ descendents can be easily worked
out; in the case $k^2=0$, the explicit forms of the $V^{(\pm)}_d$
vertices are
	\begin{eqnarray}
   \label{Vvert} V^{(+)}_d&=&-\sqrt{2}\Bigl[\xi\cdot\partial X -
	ik\wedge\xi\cdot U + k^\mu(k\wedge\xi)^\nu W_{\mu\nu} \Bigr]
	{\rm e}^{ik\cdot X} ,\nonumber\\
	V^{(-)}_d&=&+\sqrt{2}\Bigl[ik\wedge\xi\cdot\partial X -
	{\textstyle{1\over2}} \xi\cdot U -ik^\mu\xi^\nu W_{\mu\nu}
	\Bigr]{\rm e}^{ik\cdot X} .
	\end{eqnarray}
Note that in Minkowski
space-time, $(k\wedge\xi)^\mu$ is proportional to $k^\mu$ for
light-like $k^\mu$ and transverse $\xi^\mu$.  {}From (\ref{2.26}) the
highest-weight states $W^\pm_d$ have cuts in their operator products
with the currents $G^\pm$, reflecting these states' fractional
statistics on the world-sheet.  On the other hand, the $V^{(\pm)}_d$
states are world-sheet bosons since they have no cuts with $G^\pm$.
Just as with the $W_s$ state, we say that $(W^\pm_d,V^{(\pm)}_d)$ form
a fractional supermultiplet, but with conformal dimensions
$({1\over2}k^2+{1\over3},{1\over2}k^2+1)$ and world-sheet statistics
({\it fractional~, bosonic}).

\subsection{Three-point couplings and scattering amplitudes}

We will now calculate some tree-level scattering amplitudes of the
vector and scalar particles described above.  The prescription for
computing these amplitudes was worked out in Section \ref{sIII}, and is
summarized in Eqs.~(\ref{Asum}--\ref{Aselect}).  The intercept
condition (\ref{gopsc}) for the D-module physical states implies that
our vector vertex must have dimension 1/3.  $W^\pm_d$ meet this
requirement if $k^2=0$, thus describing massless vector particles.
Furthermore, the operators $\partial X_\mu$, $U_\mu$, and $W_{\mu\nu}$
appearing in $V^{(+)}_d$ are all bosonic fields on the
world-sheet---they have no cuts in their OPEs with any other
field---making them suitable for vertices in dual amplitudes by the
arguments presented in Section \ref{sIII}.  The intercept for the
S-module vertex $W_s$ describing scalar particles is not fixed by our
considerations so far.  If, for example, its intercept were 1/3, the
same as that of the D-module, then $W_s$ would describe a tachyon.  We
will mention below some considerations which may fix the S-module
intercept, but for the present discussion we will leave it arbitrary.

The simplest amplitude to calculate is the three-point coupling of two
scalar states to a vector state given by the formula (\ref{Absum})
	\begin{equation}
	 {\cal A}_{ssv} = \langle W_s(k_3)| V^{(+)}_d(k_2,\xi_2; 1)
	|W_s(k_1)\rangle,
	\end{equation}
 where we have indicated the momenta and
polarization vectors associated to each vertex.  Inserting the explicit
expressions for the vertices given in Eqs.~(\ref{2.17}) and
(\ref{Vvert}), we find
	\begin{equation}
	 {\cal A}_{ssv} = -\sqrt2 \langle k_3,0| \xi_2\cdot\left[
	\partial X(1) + i k_2\wedge U(1) - k_2\wedge W\cdot k_2 \right]
	e^{ik_2\cdot X}(1) |k_1,0\rangle.
	\end{equation}
 The $U_\mu$ and
$W_{\mu\nu}$ fields of the {\boldmath$\varphi$}\ CFT give no
contribution by Lorentz invariance and all that survives is a standard
free-boson correlator in the $X^\mu$ CFT, giving
	\begin{equation}
	 {\cal A}_{ssv} = i\sqrt2 \xi_2\cdot k_1
	\delta^3(k_1+k_2+k_3).
	\end{equation}
 It should be clear that the
calculation of any N-point function will reduce to free-field
correlators in the {\boldmath$\varphi$}\ and $X^\mu$ CFTs.

Another three-point amplitude that can be calculated in our formalism
is the coupling between one scalar and two vector particles.  It is
trivial to check that it vanishes identically, illustrating the
selection rule (\ref{Aselect}).  This implies that the scalar particle
can be consistently decoupled (at tree level) from the scattering of
vector particles, in close analogy to the way the tachyonic state in
the Neveu-Schwarz sector of the ordinary superstring can be decoupled
from scattering of the massless vector states.  In general, the
selection rule (\ref{Aselect}) allows the tree-level decoupling of
world-sheet S-module physical states from the scattering of D-module
ones, in close analogy to the GSO projection \cite{GSO} in the
Neveu-Schwarz sector of the ordinary superstring.

A less trivial amplitude is the coupling ${\cal A}_{vvv}$ of three
massless vector states.  One expects such a coupling to be
gauge-invariant since the $V^{(+)}_d$ vertices describe gauge bosons in
the transverse gauge $\xi\cdot k=0$.  Indeed, upon making a gauge
transformation $\delta\xi^\mu \sim k^\mu$, one finds $\delta V^{(+)}_d
\sim \partial ({\rm exp}\{ ik\cdot X \})$, a spurious state which
decouples by the arguments of Section \ref{sIII}.  In fact, with some
straightforward algebra using the kinematics of three massless
particles one computes explicitly
	\begin{eqnarray}
   \label{Avvv} {\cal A}_{vvv} &=& 2 \langle W^+_d(k_3,\xi_3)|
	V^{(+)}_d(k_2,\xi_2; 1) |W^-_d(k_1,\xi_1)\rangle \nonumber\\
	&=& i2\sqrt{2}\Bigl[ (k_1\cdot\xi_3) (\xi_2\cdot\xi_1) +
	(k_2\cdot\xi_1) (\xi_3\cdot\xi_2) + (k_3\cdot\xi_2)
	(\xi_1\cdot\xi_3) \nonumber\\ &&\qquad\mbox{} - (\xi_1\cdot
	k_2) (\xi_2\cdot k_3) (\xi_3\cdot k_1) \Bigr]
	\delta^3(k_1+k_2+k_3) .
	\end{eqnarray}
The first three terms are
precisely the expected Yang-Mills coupling; gauge group charges can be
introduced by Chan-Paton factors \cite{CP} in the usual way.  The last
term in (\ref{Avvv}) represents a non-linear correction to the
Yang-Mills action which is higher-order in the string tension, and
therefore is suppressed at energies far below the Planck scale. The
non-linear term also appears in the three-vector coupling in the
bosonic string, though with the opposite sign; in the superstring no such
term appears in the three-point coupling (though string correction
terms do appear in higher-point functions).

Higher-point amplitudes can also be calculated using the prescription
of the last section.  The main features of these amplitudes can be
easily understood without detailed computation.  Consider, for example,
the four-point vector particle amplitude
	\begin{equation}
	 {\cal A}_{4v} = \int_1^\infty dx \langle
	V^{(+)}_d(k_4,\xi_4)| V^{(+)}_d(k_3,\xi_3;x)
	V^{(+)}_d(k_2,\xi_2;1) \ket{V^{(+)}_d(k_1,\xi_1)},
	\end{equation}
 in the
``picture'' of Eq.~(\ref{Aone}).  Inserting the expression
(\ref{Vvert}) for $V^{(+)}_d$ leads to a sum of terms, each of which is
a product of a correlator in the {\boldmath$\varphi$}\ CFT and a
correlator in the $X_\mu$ CFT.  Now, only in the $X_\mu$ CFT
correlators is the dependence on the momenta $k_i$ non-polynomial,
entering through the exponentials as
	\begin{equation}
	\label{xcorr} \langle k_4| e^{ik_3\cdot X}(x) e^{ik_2\cdot X}(1)
	\ket{k_1}
	\end{equation}
 (perhaps with extra $\partial X^\mu$ insertions
as well).  These correlators are precisely the ones that enter into
bosonic and superstring scattering amplitudes, and give rise to
gamma-function dependence on the Mandelstam invariants similar to that
which appears in the Veneziano amplitude.  These factors result in the
extremely soft high-energy Regge behavior characteristic of string
amplitudes.  Fractional superstring amplitudes will differ from
ordinary superstring amplitudes only by the {\boldmath$\varphi$}\ CFT
correlators which are polynomial in the momenta, and so cannot change
the soft high-energy behavior of the amplitudes.  This implies, in
particular, that the cancelled propagator argument used in the last
section is justified.

It would be an interesting exercise to calculate the explicit
expression for some four-point functions in this three-dimensional
fractional superstring model.  As an example of what one could learn
from such a computation, consider the four-point correlator $\langle
W_s| V^{(+)}_d V^{(+)}_d \ket{W_s}$ of two vector and two scalar
particles.  Though our prescription for including two S-module physical
states in scattering amplitudes is not manifestly dual, the final
expression should be.  That means in practice that one could factorize
the expression in the $s$ and $t$ channels and check that the
appropriate spectra of intermediate states is recovered.  This should
place restrictions on the allowed intercepts for S-module physical
states.

A no-ghost theorem for this three-dimensional model of fractional
superstrings has been proved in Ref.~\cite{AKT}, showing that the space
of physical states has non-negative norm. Combined with the spurious
state decoupling theorem for tree scattering amplitudes shown in
Section~\ref{sIII} this implies that tree-level amplitudes in the
three-dimensional model of spin-4/3 fractional superstrings are
unitary.

Higher-point closed fractional superstring or heterotic-type scattering
amplitudes can be easily obtained by combining appropriate open string
amplitudes \cite{KLT}.  For example, in a closed string we could match
a left-moving and a right-moving version of, say, $W^+_d$ to form the
massless physical state
	\begin{eqnarray}
	 W^+_d(z,\bar z) = \xi^{\mu\nu}\Bigl[ \epsilon^+_\mu
	\overline\epsilon^+_\nu +g_{\mu\nu} (k\cdot\epsilon^+)
	(k\cdot\overline\epsilon^+)\nonumber\\ + \epsilon^+_\mu
	(ik\wedge\overline\epsilon^+)_\nu +(ik\wedge\epsilon^+)_\mu
	\overline\epsilon^+_\nu \Bigr] e^{ik\cdot X}
	\end{eqnarray}
with
$k_\mu\xi^{\mu\nu}=k_\nu\xi^{\mu\nu}=0$.  The symmetric-traceless,
antisymmetric, and trace parts of $\xi^{\mu\nu}$ will then describe the
graviton, the antisymmetric tensor field, and the dilaton (in covariant
gauge), respectively, just as in bosonic strings and ordinary
superstrings.

\section{\label{sVII}Discussion and outlook}

In this paper we have shown how to construct tree-level scattering
amplitudes for the spin-4/3 fractional superstring which are dual and
obey spurious state decoupling.  We have illustrated these properties
in an explicit three-dimensional model of the spin-4/3 fractional
superstring, and found that it has a sensible space-time spectrum
including gauge bosons and a graviton (for closed strings).  Tree-level
unitarity follows from the spurious state decoupling property once a
``no-ghost'' theorem for the physical state spectrum in a given
representation is proven.  A no-ghost theorem has been proved in
Ref.~\cite{AKT} for the three-dimensional model discussed in this
paper.  Space-time fermion states for the $c=5$ representation of the
spin-4/3 algebra are constructed in Ref.\ \cite{tree2}, where general
(representation-independent) spurious state decoupling arguments are
also presented for scattering amplitudes involving twisted-sector
states.

The tree-level considerations of this paper and Ref.\ \cite{tree2}
leave us with a certain amount of arbitrariness in constructing
spin-4/3 fractional superstrings.  In particular, we are free to
include or not the world-sheet S-module untwisted-sector states; we can
couple left- and right-moving theories at will on the world-sheet in
type II and heterotic constructions; and the choice of CFT
representation of the spin-4/3 FSC algebra is presumably constrained by
tree unitarity only to have central charge less than or equal to its
critical value $c=10$.  The inclusion of string loop amplitudes should
remove much of this arbitrariness.  As is the case with the bosonic and
superstrings, one expects that loop amplitudes will only be consistent
at the critical central charge, and modular invariance will determine
which left- and right-moving sectors, and at which values of their
intercepts, can be consistently coupled together.

The main difficulty in constructing a critical ($c=10$) representation
of the FSC algebra is its non-linearity discussed in Section
\ref{sII}:  the tensor product of two representations of this algebra
is not itself a representation.  In particular, the tensor product of
two copies of the $c=5$ representation described above will not make a
$c=10$ representation of the FSC algebra.  For certain representaions,
one can, however, construct higher-$c$ representations from a given
representation by turning on a background charge for one of the
$X^\mu(z)$ coordinate boson fields, corresponding to turning on a
linear dilaton background in space-time.  Also, a set of
representations constructed from free bosons have been found, all with
central charges $c\le8$.  These representations are briefly described
in \ref{appC}.  It may be that some generalization of these
constructions will yield $c>8$ (and in particular $c=10$)
representations.

Once given a $c=10$ representation of the FSC algebra, one can imagine
``sewing'' tree amplitudes in the old covariant formalism described
above to form one-loop amplitudes by a suitable generalization of the
sewing procedure for the bosonic string \cite{BrO}.  Such an amplitude
would not only have to be unitary, but also modular invariant.  The
construction of a consistent one loop amplitude is a crucial test of
the existence of the spin-4/3 fractional superstring theory.

At higher loops it seems likely that a clearer understanding of the
``fractional moduli'' describing the sewing of tree amplitudes will be
necessary.  This is essentially the question of determining the local
world-sheet symmetry underlying the FSC constraint algebra.  Though the
form of the FSC algebra provides a rigid guide to such a symmetry, its
identification remains an open question.  One approach to this problem
is to construct a world-sheet ghost system with a nilpotent BRST charge
whose cohomology reproduces the physical state conditions analyzed in
this paper.  The BRST charge would be expected to have the form $Q =
cT_m + \gamma^+ G^-_m + \gamma^- G^+_m +\ldots$ where $T_m$ and
$G^\pm_m$ are the ``matter'' FSC currents, $c$ is the dimension-$(-1)$
reparametrization ghost, and $\gamma^\pm$ are dimension-$(-{1\over3})$
fractional superghost fields.  No such ghost system and BRST charge
have been constructed.  Another possibility is that the BRST ghosts of
the fractional superstring and the matter fields are inherently
coupled.  In this case one should seek a $c=0$ representation of the
FSC algebra that contains both the matter and the ghost fields and
permits the construction of a nilpotent BRST charge.

In this paper, we considered the fractional superstring based on the
spin-4/3 FSC algebra. This world-sheet algebra is associated with the
$su(2)_4$ Wess-Zumino-Witten (WZW) model as explained in \ref{appC}.
In general, one can construct fractional algebras associated in the
same way to WZW models based on any Lie algebra \cite{KMQ,ALT,AGT}.
For example, the algebra based on $su(2)_1$ is simply the Virasoro
algebra, and its associated string is the bosonic string; associated
with $su(2)_2$ is the super-Virasoro algebra which underlies the
ordinary superstring.  These are special examples in that the resulting
algebras are local on the world-sheet.  Other local algebras underly
the $W$-strings associated with any level-one WZW model.  Given the
results of this paper, it is natural to speculate that there exist
strings corresponding to the non-local algebras associated with WZW
models at arbitrary levels.

The generic such fractional string, however, will be technically
more difficult to work with than the spin-4/3 fractional string.
The main reason is that the general fractional chiral algebra does
not admit a splitting into abelianly braided currents as the
spin-4/3 FSC algebra did.  This splitting was the main technical
crutch that allowed us to understand the properties of the FSC
modules in a representation-independent way.  To deal similarly
with an inherently non-abelianly braided current algebra will
require a more thorough understanding of the braiding properties
of their currents and the development of the conformal field theory
techniques needed to derive their generalized commutation relations.

Two particularly simple series of fractional superconformal algebras
are those based on the $so(N)_2$ models for arbitrary $N$, and those
based on the $su(2)_K$ models for arbitrary $K$.  Since conformally,
$su(2)_4=so(3)_2$, the former series includes the spin-4/3 fractional
superstring as a special case.  The merit of this series is that all
the resulting fractional world-sheet algebras are abelianly braided.
Also they clearly have representations with global $so(N)$ symmetry
groups; however it is not clear how to construct representations with
$N$-dimensional Poincar\'{e} invariance.  It is interesting to note
that, since $so(4)=so(3)\otimes so(3)$, the world-sheet symmetry
algebra corresponding to $so(4)_2$ is simply two copies of the spin-4/3
algebra; however, the coordinate bosons coupled to the $so(3)_2$ model
in our $c=5$ representation do not transform in the vector
representation of $so(4)$, and so cannot give a flat space-time
interpretation.

The representation theory of the other simple series of algebras based
on the $su(2)_K$ models, though non-abelianly braided in general, have
been more intensively studied \cite{AGT}.  Since $su(2)_K=so(3)_{K/2}$
all these models (trivially) have representations with $so(2,1)$
Lorentz symmetry.  Whether there are any representations in which this
Lorentz symmetry can be extended to the Poincar\'{e} symmetry of three-
(or higher-) dimensional space-time is an open problem.  One indication
that such representations really may exist comes from the
modular-invariant fractional superstring partition functions proposed in
Ref.~\cite{AT}.  Although the precise connection between these
partition functions and fractional superstrings defined by a fractional
superconformal algebra as discussed above is not clear, there are many
suggestive points of contact; indeed, the construction of the explicit
three-dimensional $c=5$ representation of the spin-4/3 algebra was
originally motivated by consideration of the ``internal projection''
appearing in these partition functions \cite{ADT,AD}.  Some hints of
the space-time structure of the critical $su(2)_K$ fractional
superstrings have been gleaned from fractional superstring partition
functions \cite{AT,ADT}.  For example, the low-energy physics of these
strings is believed to describe supergravity in six and four dimensions
for $K=4$ and $8$, respectively.  If this is true, then the critical
spin-4/3 fractional superstring should have a six flat space-time
dimensional representation.  An interesting question in connection with
these new strings is whether their fractional world-sheet structures
``translate'' into some novel symmetries or physics in space-time.  In
this connection, there are some suggestive hints from the fractional
superstring partition functions \cite{AD,DT}.

\vspace{2.5ex}
\begin{flushleft}
\large\bf Acknowledgements
\end{flushleft}

It is a pleasure to thank K. Dienes, J. Grochocinski, Z. Kakushadze
and A. LeClair for useful discussions and comments.  Much of this work
was carried out at Cornell University and was supported in part by the
National Science Foundation.  P.C.A. would also like to thank the Center
for Theoretical Physics at M.I.T., and B.~Zwiebach in particular, for
their hospitality.  The work of P.C.A. is supported by NSF grant
PHY92-45317 and by the Ambrose Monell Foundation.

\setcounter{section}{0}
\Appendix{\label{appA}Untwisted sector of $so(2,1)_2$}

We add standard cocycles to the free boson theory compactified on the
$su(3)$ root lattice considered in section \ref{sII}.  Following the
notation of that section, define the vertex operators $V_{\bf m}$ as
	\begin{equation}
	\label{A1} V_{\bf m} = c({\bf m}) :e^{i{\bf m}\cdot
	\mbox{\boldmath$\varphi$}}:,
	\end{equation}
where the colons denote normal ordering with respect to the conformal
vacuum.  The cocycles $c({\bf m})$ can be chosen to obey the properties
\cite{FK}
	\begin{eqnarray}
   \label{A2} c({\bf m})c({\bf n}) &=& c({\bf m}+{\bf n})\nonumber\\
	c({\bf m}) e^{i{\bf n}\cdot
	\mbox{\boldmath$\varphi$}} &=& (-1)^{m_1n_2} e^{i{\bf
	n}\cdot\mbox{\boldmath$\varphi$}} c({\bf m})\nonumber\\
	{}[c({\bf m})]^\dagger &=&
	c(-{\bf m})\nonumber\\ c({\bf 0}) &=& 1.
	\end{eqnarray}
These properties
imply, in particular, that
	\begin{equation}
	\label{A3} (V_{\bf m})^\dagger = (-1)^{m_1m_2}V_{-{\bf m}} .
	\end{equation}
Using these definitions and the free field operator products
(\ref{2.6}), the basic vertex operator product expansion
	\begin{equation}
	\label{Abope} V_{\bf m}(z) V_{\bf n}(w) = (-1)^{m_2n_1}(z-w)^{\bf
	m\cdot n} V_{\bf m+n}(w) + \ldots
	\end{equation}
 can be derived.

Alternatively, the hilbert space can be explicitly constructed in terms
of the modes of the $\mbox{\boldmath$\varphi$}(z)$ fields, defined by
the expansion
	\begin{equation}
	 \varphi^j(z)=\phi^j-i p^j\ln(z) +i\sum_{n\neq0}{1\over
	n}\alpha^j_n z^{-n} ,
	\end{equation}
 and satisfing the commutation
relations $[\phi^i, p^j]=ig^{ij}$, $[\alpha^i_n,\alpha^j_m]= n
g^{ij}\delta_{m+n}$, and the hermiticity assignments
$(\phi^j)^\dagger=\phi^j$, $(p^j)^\dagger=p^j$, and
$(\alpha^j_n)^\dagger=\alpha^j_{-n}$.  The cocycles can be explicitly
realized \cite{KLT} by $c({\bf m})=(-1)^{m_1p_2}$, where
$p_2=g_{2j}p^j$ is a component of the momentum zero-mode of
$\mbox{\boldmath$\varphi$}(z)$.  In terms of the modes, the
normal-ordered vertex operators can be written
	\begin{eqnarray}
   \label{novtx} V_{\bf m}(z)&=&(-1)^{m_1p_2} {\rm exp}\left\{i{\bf
	m}\cdot\mbox{\boldmath$\phi$}\right\} {\rm exp}\left\{{\bf
	m\cdot p}{\rm ln}(z)\right\} \nonumber\\ &&\quad\times {\rm
	exp}\left\{-\sum_{n<0}{1\over n}{\bf
	m}\cdot\mbox{\boldmath$\alpha$}_n z^{-n}\right\} {\rm
	exp}\left\{-\sum_{n>0}{1\over n}{\bf
	m}\cdot\mbox{\boldmath$\alpha$}_n z^{-n}\right\} .
	\end{eqnarray}
The
basic operator product expansion (\ref{Abope}) follows easily.

All fields in the {\boldmath$\varphi$}\ CFT can be organized in
$so(2,1)$ representations.  All the Virasoro primary fields in the
{\boldmath$\varphi$}\ CFT up to dimension 4/3 are
	\begin{eqnarray}
   \label{Acovflds} h=1/3:\hfill\qquad\qquad \epsilon^+_\mu &=&
	\left( V_{(-1,-1)}, V_{(1,0)}  , V_{(0,1)} \right) \nonumber\\
	\epsilon^-_\mu &=& \left( V_{(1,1)}  , V_{(-1,0)} , V_{(0,-1)}
	\right) \nonumber\\ & & \nonumber\\
	h=\ 1\ \,:\hfill\qquad\qquad U_\mu &=& \left(
	V_{(1,-1)}+V_{(-1,1)} \, , \, V_{(1,2)}+V_{(-1,-2)} \, , \,
	V_{(-2,-1)}+V_{(2,1)} \right) \nonumber\\ W_{\mu\nu} &=&
	{1\over2} \pmatrix{ 2i\partial\varphi_{(1,1)} &
	V_{(2,1)}-V_{(-2,-1)} & V_{(-1,-2)}-V_{(1,2)} \cr &
	2i\partial\varphi_{(1,0)} & V_{(1,-1)}-V_{(-1,1)} \cr & &
	2i\partial\varphi_{(0,1)} \cr} \nonumber\\ & & \nonumber\\
	h=4/3:\hfill\qquad\qquad s^+ &=& {1\over3} \left[
	V_{(2,2)}+V_{(-2,0)}+V_{(0,-2)} \right] \\ s^- &=& {1\over3}
	\left[ V_{(-2,-2)}+V_{(2,0)}+V_{(0,2)} \right] \nonumber\\
	t^+_{\mu\nu} &=& \pmatrix{ -V_{(2,2)}+s^+ &
	-{i\over2}\partial\varphi_{(2,1)}V_{(0,1)} &
	{i\over2}\partial\varphi_{(1,2)}V_{(1,0)}\cr & V_{(-2,0)}-s^+ &
	{i\over2}\partial\varphi_{(-1,1)}V_{(-1,-1)}\cr & &
	V_{(0,-2)}-s^+ \cr}\nonumber\\ t^-_{\mu\nu} &=& \pmatrix{
	-V_{(-2,-2)}+s^- & {i\over2}\partial\varphi_{(2,1)}V_{(0,-1)} &
	-{i\over2}\partial\varphi_{(1,2)}V_{(-1,0)}\cr & V_{(2,0)}-s^-
	& -{i\over2}\partial\varphi_{(-1,1)}V_{(1,1)}\cr & &
	V_{(0,2)}-s^- \cr}\nonumber
	\end{eqnarray}
Here we have defined the
combination $\varphi_{\bf m} = {\bf m}\cdot\mbox{\boldmath$\varphi$}$,
so that, for example, $\partial\varphi_{(2,1)} = 2\partial\varphi^1 +
\partial\varphi^2$.  We have also only written half of the entries for
the spin-2 fields $W_{\mu\nu}$ and $t^\pm_{\mu\nu}$, since they are
symmetric-traceless tensors.

{}From (\ref{A3}) it follows that hermitian conjugation is accompanied
by lowering (raising) upper (lower) space-time indices.  For example,
$(\epsilon^{+\mu})^\dagger=\epsilon^-_\mu$, and
$(W_{\mu\nu})^\dagger=W^{\mu\nu}$.

The vertex operator OPEs can be worked out using the free field
operator products (\ref{2.6}) and the vertex OPEs (\ref{Abope}).  The
results for some leading terms are listed below.  For ease of writing,
all the operator products are of the form $A(z)B(0)$, the right-hand
sides of the OPEs are all evaluated at $0$, and the
dependence of the fields on their arguments is suppressed.
	\begin{eqnarray}
   \label{eeope} \epsilon^\pm_\mu \epsilon^\pm_\nu &=& z^{-1/3}
	{\varepsilon_{\mu\nu}}^\rho \epsilon^\mp_\rho + z^{2/3} \left(
	{\textstyle{1\over2}} {\varepsilon_{\mu\nu}}^\rho
	\partial\epsilon^\mp_\rho + g_{\mu\nu} s^\mp +
	t^\mp_{\mu\nu}\right) \nonumber\\ \epsilon^\pm_\mu
	\epsilon^\mp_\nu &=& z^{-2/3} g_{\mu\nu} + z^{1/3} \left(
	{\textstyle{1\over2}} \varepsilon_{\mu\nu\rho} U^\rho \pm
	W_{\mu\nu}\right) \nonumber\\ & & \mbox{} + z^{4/3} \left(
	{\textstyle{1\over3}} g_{\mu\nu} T_\varphi +
	{\textstyle{1\over4}} \varepsilon_{\mu\nu\rho} \partial U^\rho
	\pm {\textstyle{1\over2}} \partial W_{\mu\nu} + H_{\mu\nu}
	+\varepsilon_{\mu\nu\rho}F^\rho\right)
	\end{eqnarray}

	\begin{eqnarray}
   \label{esope} \epsilon^\pm_\mu s^\pm &=& z^{-4/3}
	{\textstyle{1\over3}} \epsilon^\mp_\mu - z^{-1/3}
	{\textstyle{1\over3}} \partial \epsilon^\mp_\mu \nonumber\\
	s^\pm \epsilon^\pm_\mu &=& z^{-4/3} {\textstyle{1\over3}}
	\epsilon^\mp_\mu + z^{-1/3} {\textstyle{2\over3}} \partial
	\epsilon^\mp_\mu \nonumber\\ \epsilon^\pm_\mu s^\mp &=&
	z^{-2/3} {\textstyle{1\over3}} U_\mu -
	z^{1/3}{\textstyle{1\over3}}F_\mu \nonumber\\ s^\pm
	\epsilon^\mp_\mu &=& z^{-2/3} {\textstyle{1\over3}} U_\mu +
	z^{1/3} {\textstyle{1\over3}} \left(\partial U_\mu +
	F_\mu\right)
	\end{eqnarray}

	\begin{eqnarray}
	 s^\pm s^\pm &=& z^{-4/3} {\textstyle{2\over3}} s^\mp +
	z^{-1/3} {\textstyle{1\over3}} \partial s^\mp \nonumber\\ s^\pm
	s^\mp &=& z^{-8/3} {\textstyle{1\over3}} + z^{-2/3}
	{\textstyle{4\over9}} T_\varphi
	\end{eqnarray}

	\begin{eqnarray}
	 \epsilon^\pm_\mu U_\nu &=& z^{-1} {\varepsilon_{\mu\nu}}^\rho
	\epsilon^\pm_\rho - {\textstyle{1\over2}}
	{\varepsilon_{\mu\nu}}^\rho \partial\epsilon^\pm_\rho +
	2g_{\mu\nu} s^\pm - t^\pm_{\mu\nu} \nonumber \\ U_\mu
	\epsilon^\pm_\nu  &=& z^{-1} {\varepsilon_{\mu\nu}}^\rho
	\epsilon^\pm_\rho + {\textstyle{3\over2}}
	{\varepsilon_{\mu\nu}}^\rho \partial\epsilon^\pm_\rho +
	2g_{\mu\nu} s^\pm - t^\pm_{\mu\nu} \nonumber\\ s^\pm U_\mu &=&
	z^{-2} {\textstyle{2\over3}} \epsilon^\pm_\mu + z^{-1}
	{\textstyle{2\over3}} \partial\epsilon^\pm_\mu \nonumber\\
	U_\mu s^\pm &=& z^{-2} {\textstyle{2\over3}} \epsilon^\pm_\mu +
	{\cal O}(z^0)
	\end{eqnarray}

	\begin{eqnarray}
	 \epsilon^\pm_\rho W_{\mu\nu} &=& \mp z^{-1}
	{\textstyle{1\over2}} {\delta_{\mu\nu\rho}}^\sigma \left(
	\epsilon^\pm_\sigma - z {\textstyle{1\over2}}
	\partial\epsilon^\pm_\sigma \right) \mp
	{\textstyle{1\over2}} \left( {\varepsilon_{\rho\mu}}^\sigma
	t^\pm_{\sigma\nu} + {\varepsilon_{\rho\nu}}^\sigma
	t^\pm_{\sigma\mu} \right) \nonumber\\ W_{\mu\nu}
	\epsilon^\pm_\rho &=& \pm z^{-1} {\textstyle{1\over2}}
	{\delta_{\mu\nu\rho}}^\sigma \left( \epsilon^\pm_\sigma + z
	{\textstyle{3\over2}} \partial\epsilon^\pm_\sigma \right)
	\pm {\textstyle{1\over2}} \left( {\varepsilon_{\rho\mu}}^\sigma
	t^\pm_{\sigma\nu} + {\varepsilon_{\rho\nu}}^\sigma
	t^\pm_{\sigma\mu} \right) \nonumber\\ s^\pm W_{\mu\nu} &=& \pm
	z^{-1} {\textstyle{2\over3}} t^\pm_{\mu\nu} \nonumber\\
	W_{\mu\nu} s^\pm &=& \mp z^{-1} {\textstyle{2\over3}}
	t^\pm_{\mu\nu}
	\end{eqnarray}
where $g_{\mu\nu}$ is the Minkowski metric
in three dimensions with signature $(-++)$, $\varepsilon_{\mu\nu\rho}$
is the antisymmetric tensor in three dimensions normalized by
$\varepsilon_{012}=-1$, obeying $\varepsilon_{\mu\nu\rho}
{\varepsilon^\mu}_{\alpha\beta} = -g_{\nu\alpha} g_{\rho\beta} +
g_{\nu\beta} g_{\rho\alpha}$, and we have defined
$\delta_{\mu\nu\rho\sigma} = g_{\mu\rho} g_{\nu\sigma} + g_{\mu\sigma}
g_{\nu\rho} - {2\over3} g_{\mu\nu} g_{\rho\sigma}$.  The fields $F_\mu$
and symmetric-traceless $H_{\mu\nu}$ appearing in  (\ref{eeope}) and
(\ref{esope}) are dimension-2 Virasoro primary fields.  We can take
(\ref{eeope}) as their definition.

\Appendix{\label{appC}Other known spin-4/3 FSC representations}

The representation theory of the spin-4/3 FSC algebra is related to the
$su(2)_4$ Wess-Zumino-Witten (WZW) model in the same way as the
Virasoro algebra representation theory is related to the $su(2)_1$
model.  In particular, the FSC algebra has a series of unitary minimal
representations realized by the $su(2)_4\otimes su(2)_L/su(2)_{4+L}$
coset models with central charges
	\begin{equation}
	\label{coset} c = 2-{24\over(L+2)(L+6)}\qquad {\rm
	for}\ \ L=1,2,\ldots ,
	\end{equation}
 accumulating at the $c=2$ $su(2)_4$
WZW model. The FSC algebra can be realized in this model as follows.
Let $J^a(z) = \sum_n J^a_n z^{-n-1}$ denote the $su(2)_4$ Ka\v{c}-Moody
currents, $\Phi^a(z)$ the dimension-1/3 chiral primary field in the
adjoint representation, and $q_{ab}$ the $su(2)$ Killing form.  The FSC
current is \cite{KMQ,ALT}
	\begin{equation}
	 G^+(z) + G^-(z) = \sum_{a,b}q_{ab}J^a_{-1}\Phi^b(z) .
	\end{equation}
The $su(2)_4$ theory can be bosonized in terms of one free boson and a
$\IZ_4$ parafermion theory \cite{ZFpf}, which is itself equivalent to a
single compactified boson.  Thus the $c=2$ representation can be
written in terms of two free bosons, $X$ and $\rho$, satisfying
$X(z)X(w) = -{\rm ln}(z-w)$ and $\rho(z)\rho(w) = -{1\over6}{\rm
ln}(z-w)$, with $\rho$ compactified on the unit circle $\rho =
\rho+2\pi$.  The FSC algebra currents are given by \cite{ALyT}
	\begin{equation}
	 G^\pm = {i\over\sqrt2}e^{\pm2i\rho}\partial
	X+{1\over2}e^{\mp4i\rho} .
	\end{equation}
 The coset models can be
realized in terms of this bosonized theory by turning on a background
charge for the $X$ boson \cite{KMQ}, $T_X = -{1\over2}(\partial X
\partial X + i Q \partial^2 X )$, so that the total central charge of
the representation becomes $c = 2-3Q^2$.  The expression for the FSC
current with background charge is given in Ref.~\cite{ALT}.

The $su(2)_4$ WZW model is equivalent to the $so(3)_2$ model, whose
two-boson construction was explained in Section \ref{sIV}.  This
equivalence suggests other free field representations of the spin-4/3
FSC algebra.  In particular, an inequivalent $c=2$ representation can
be realized in terms of the two bosons of the $so(3)_2$ model.  The
split algebra currents in this representation are simply \cite{GoS}
	\begin{equation}
	 G^\pm = {3\over2}s^\pm ,
	\end{equation}
 where the $s^\pm$ fields are
defined in \ref{appA}.  Other free boson representations can be formed
by taking various tensor products of free uncompactified bosons $X^\mu$
and copies of the $so(3)_2$ WZW model.

A $c=4$ representation is $so(3)_2\otimes so(3)_2$ with the split
algebra currents given by
	\begin{equation}
	 G^\pm = {1\over\sqrt6}\left(e^{\pm2\pi i/9} U^\mu \otimes
	\epsilon^\pm_\mu +3 e^{\mp\pi i/9} \cdot {\bf 1} \otimes
	s^\pm\right) .
	\end{equation}
 where ${\bf 1}$ denotes the identity
operator.  This representation is related to Goddard and Schwimmer's
construction \cite{GoS} of the subset of the spin-4/3 FSC algebra
minimal models with $L=2K$ in (\ref{coset}) in terms of an
$so(3)_K\otimes so(3)_2$ theory with a $U^\mu \otimes U_\mu$ term added
to the stress-energy tensor.

The $c=5$ representation discussed in detail in this paper is a tensor
product of one $so(3)_2$ with three free bosons $X^\mu$.  We repeat
here the resulting form of the split algebra currents:
	\begin{equation}
	 G^\pm = {1\over\sqrt2}\left(\pm\partial
	X^\mu\epsilon^\pm_\mu -{3\over2}s^\pm\right) .
	\end{equation}
 It is a
non-trivial fact that when a background charge is turned on for the
$X^\mu$ fields in this representation, the form of the fractional current
can be modified in such a way as to still satisfy the FSC algebra \cite{KT}.

A $c=6$ representation is the three-fold tensor product $so(3)_2\otimes
so(3)_2\otimes so(3)_2$ with split algebra currents given by
	\begin{equation}
	 G^\pm = {1\over\sqrt6}\left(e^{\pm\pi i/4}U^\mu\otimes{\bf
	1}\otimes \epsilon^\pm_\mu + e^{\mp\pi i/4}{\bf 1}\otimes
	U^\mu\otimes \epsilon^\pm_\mu +{3\over\sqrt2}{\bf 1}\otimes{\bf
	1}\otimes s^\pm\right) .
	\end{equation}

A $c=7$ representation consists of two copies of $so(3)_2$ and one set
of three free bosons $X^\mu$ with
	\begin{equation}
	 G^\pm = {1\over\sqrt6}\left(\pm\sqrt3\partial
	X^\mu\cdot{\bf 1}\otimes \epsilon^\pm_\mu -
	U^\mu\otimes\epsilon^\pm_\mu -{3\over2}{\bf 1}\otimes
	s^\pm\right) .
	\end{equation}
 It is interesting to note that this
representation has a three-dimensional Poincar\'{e} symmetry, and so,
like the $c=5$ representation discussed in this paper, is a suitable
model for constructing spin-4/3 fractional string tree-level scattering
amplitudes.

A $c=8$ representation consists of one $so(3)_2$ and two sets
of three free bosons, $X^\mu$ and $Y^\mu$.  Its split algebra currents
have the especially simple form
	\begin{equation}
	 G^\pm = \pm{1\over\sqrt2}\partial\left(X^\mu\pm
	iY^\mu\right) \epsilon^\pm_\mu .
	\end{equation}
 Note that at $c=8$ the
FSC structure constants vanishe: $\lambda^\pm=0$.  Although this
representation has a six-dimensional global translation group, its
largest global ``rotation'' group is only $su(3)$, and thus cannot
describe string scattering in six dimensions.  Furthermore, since the
uncompactified boson fields appear only in the complex combinations
$X^\mu\pm iY^\mu$, there are effectively three timelike directions in
this representation, which thus has no hope of giving rise to unitary
string scattering amplitudes.

No $c>8$ free field representations ({\it i.e.}, with no background
charges) are known for the spin-4/3 FSC algebra.

Finally, note that the hermiticity properties of the currents
are constrained by the form of the FSC algebra.
Assuming that $G^+$ and $G^-$ are related by some
hermiticity relations (which by no means has to be the
case), it is not hard to show that, up to
rescalings of the currents, the algebra (\ref{fsca}) admits
four inequivalent hermiticity assignments:
\begin{equation}
	  \begin{array}{rlll}
  (i):  & \quad(G^+)^\dagger=G^-  & \quad(G^-)^\dagger=G^+
   & \quad\lambda^+=\lambda^-~,\\
  (ii): & \quad(G^+)^\dagger=G^+    & \quad(G^-)^\dagger=G^-
   & \quad\lambda^+=\lambda^-~,\\
  (iii):& \quad(G^+)^\dagger=-G^- & \quad(G^-)^\dagger=-G^+
   & \quad\lambda^+=-\lambda^-~,\\
  (iv): & \quad(G^+)^\dagger=G^+    & \quad(G^-)^\dagger=G^-
   & \quad\lambda^+=-\lambda^-~,
  \end{array}
	\end{equation}
where in all cases $\lambda^+$ can be taken to be a postive real
number.  Note that, by (\ref{lamb}), $\lambda^+\lambda^-$ changes sign
at $c=8$, so the hermiticity assignments $(i)$ and $(ii)$ apply only
when $c\le8$, while the assignments $(iii)$ and $(iv)$ are allowed only
for $c\ge8$.

For all the hermiticity assignments one can construct the hermitian
current $G\equiv G^++{\rm sign}(8-c)G^-$  which satisfies
for $c>8$, $GG\sim -1+\ldots$. This shows that, because $G$ is
hermitian, such $c>8$ representations of the FSC algebra are necessarily
non-unitary.  As mentioned in Section \ref{sI}, the critical central
charge of the spin-4/3 fractional superstring is $c=10$, and thus any
critical representation of the FSC algebra with simple hermiticity
properties for the fractional currents will be non-unitary.  This,
of course, is perfectly consistent for strings describing propagation
in Minkowski space-times; however, it is different from what occurs in
bosonic and ordinary superstrings where there is no such automatic
requirement of nonunitarity at the critical central charge.


\end{document}

\bye